\documentclass[final,5p,times,twocolumn]{elsarticle}

\usepackage{graphicx}

\usepackage{amssymb}

\usepackage{psfrag}

\journal{Nuclear Instruments and Methods in Physics Research, A}

\begin{document}

\begin{frontmatter}



\title{The convergence of EAS radio emission models and a detailed comparison of REAS3 and MGMR simulations}


\author[KITIK]{T. Huege\corref{cor}}
\author[KITEKP]{M. Ludwig}
\author[KVI]{O. Scholten}
\author[KVI]{K.D. de Vries}

\address[KITIK]{Karlsruher Institut f\"ur Technologie, Institut f\"ur Kernphysik, Campus Nord, 76021 Karlsruhe, Germany}
\address[KITEKP]{Karlsruher Institut f\"ur Technologie, Institut f\"ur Experimentelle Kernphysik, Campus S\"ud, 76128 Karlsruhe, Germany}
\address[KVI]{Kernfysisch Versneller Instituut, Zernikelaan 25, 9747 AA Groningen, The Netherlands}

\cortext[cor]{Corresponding author: Tim Huege $<$tim.huege@kit.edu$>$}

\begin{abstract}
Over the previous decade, many approaches for the modelling of radio emission from cosmic ray air showers have been developed. However, there remained significant deviations between the models, reaching from important qualitative differences such as \emph{unipolar} versus \emph{bipolar} pulses to large variations in the predicted absolute amplitudes of up to factors of 20. Only recently, it has been realized that in the many models predicting \emph{unipolar} pulses, a radio emission contribution due to the time-variation of the number of charged particles or, equivalently, the acceleration of the particles at the beginning and the end of their trajectories, had not been taken into account. We discuss here the nature of the underlying problem and demonstrate that by including the missing contribution in REAS3, the discrepancies are reconciled. Furthermore, we show a direct comparison of REAS3 and MGMR simulations for a set of prototype showers. The results of these two completely independent and very different modelling approaches show a good level of agreement except for regions of parameter space where differences in the underlying air shower model become important. This is the first time that two radio emission models show such close concordance, illustrating that the modelling of radio emission from extensive air showers has indeed made a true breakthrough.
\end{abstract}


\end{frontmatter}


\section{Introduction}

Due to renewed interest in radio detection of cosmic ray air showers, the need for reliable and detailed models of the radio emission process has increased continuously over recent years. Although the emission process had already been investigated extensively in the early days of air shower radio detection, it became clear that renewed modelling efforts were necessary. Consequently, modelling of the radio emission of EAS was restarted by a number of different groups as early as 2001. (A review on the evolution of the modern radio emission approaches can be found in \citep{HuegeArena2008}.)

When more and more models appeared on the market, it became obvious that there were contradictions between them. In particular, two classes of models could be identified. On the one hand there were models predicting \emph{unipolar} radio pulses, the frequency spectra of which levelled off at a constant value towards very low frequencies. Most strikingly, this was the case for most of the models working in the time-domain. On the other hand, there were models predicting \emph{bipolar} pulses, the frequency spectra of which fall to zero for very low frequencies. All frequency-domain approaches had this characteristic feature.

\begin{figure*}[htb]
\centering
\includegraphics[width=0.8\textwidth]{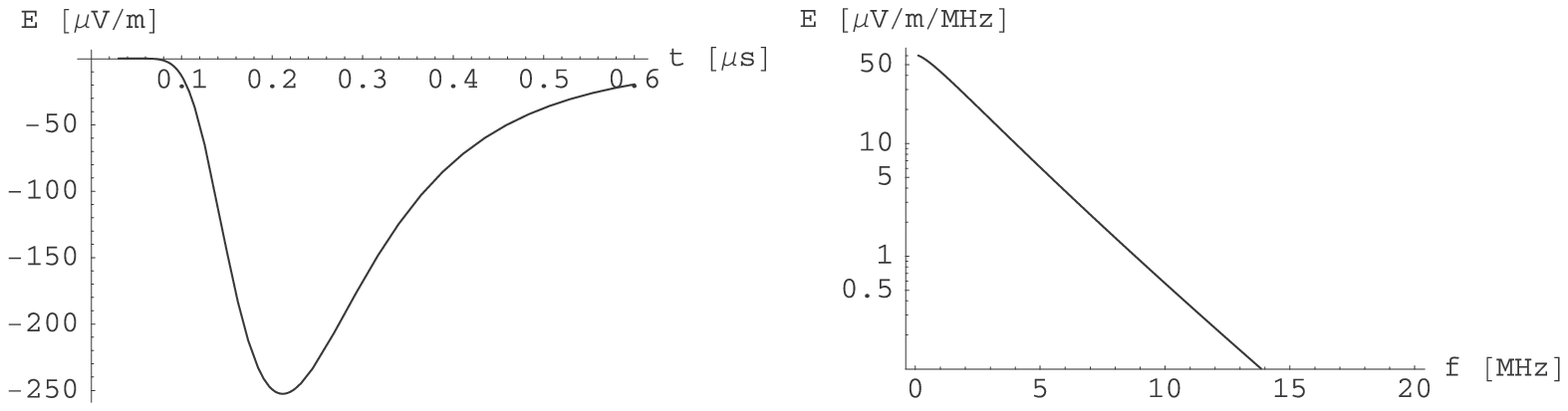}\\
\vspace{0.5cm}
\includegraphics[width=0.8\textwidth]{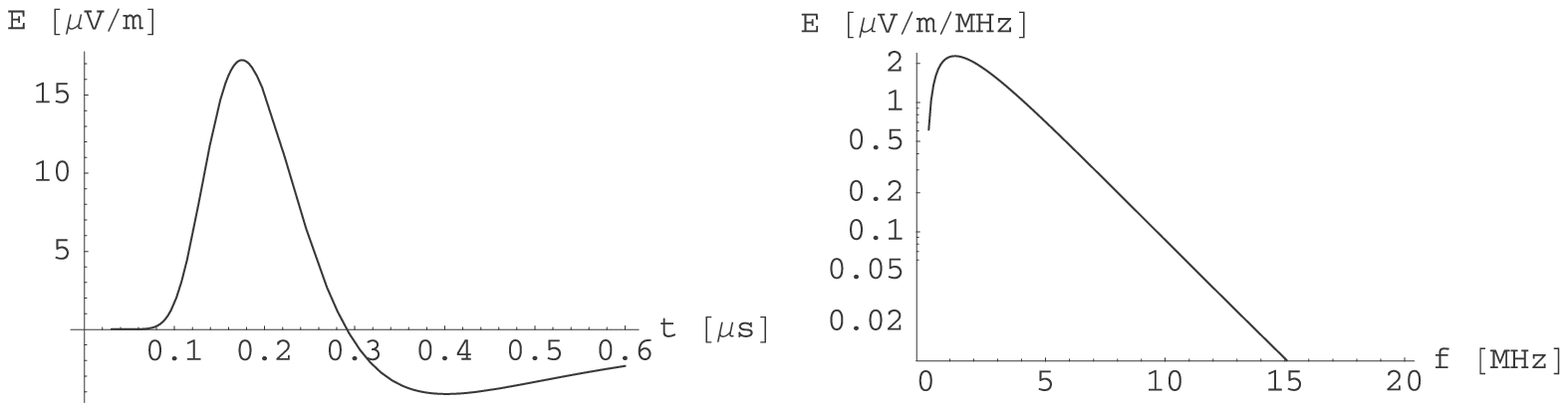}
\caption{\label{fig:gousset}Time pulses (left) and frequency spectra (right) calculated with the large impact parameter analytical radio emission model of Gousset et al.\ \citep{GoussetLamblinValcares2009} for the acceleration component (top, \emph{unipolar}) and the current component (bottom, \emph{bipolar}).}
\end{figure*}

These issues were first discussed in detail during a workshop dedicated to radio emission theory at Forschungszentrum Karlsruhe \citep{RadioTheoryMeeting2007}. In an article by Gousset, Lamblin \& Valcares \citep{GoussetLamblinValcares2009}, a direct comparison was made between a description predicting \emph{bipolar} pulses and a formulation producing \emph{unipolar} pulses (see Fig.\ \ref{fig:gousset}). The resolution of the contradiction, however, was not yet found.

From the fact that the source of the radio emission exists only over a finite time in a finite region of space, however, one can argue that the zero-frequency component of the emission (which corresponds to an infinite time-scale), can contain no power \citep{ScholtenWernerARENA2008}. Therefore, the models with \emph{unipolar} pulses, which have frequency spectra leveling off at a constant value at frequency zero, had to be suffering from some problem. The resolution of this problem has been found only recently.

In this article, we shortly review the models producing \emph{unipolar} and \emph{bipolar} pulses before explaining what causes the discrepancy and how the affected time-domain models can be corrected. Finally, we compare two very different and completely independent models with correct implementations, namely the REAS3 \citep{LudwigHuege2010,LudwigHuegeARENA2010} and MGMR \citep{ScholtenWernerRusydi2008,DeVriesARENA2010} models. We demonstrate that for the first time, two models show a good agreement in the predicted radio emission features. This marks a major milestone in the modelling of radio emission from extensive air showers.



\section{Modern modelling approaches}

In this section, we provide an (incomplete) overview of modern modelling approaches for radio emission from extensive air showers and demonstrate how they can be grouped into models predicting \emph{unipolar} pulses and models predicting \emph{bipolar} pulses. For models we could not mention here, we kindly refer the reader to the review by Huege \citep{HuegeArena2008}.

\subsection{Models with \emph{unipolar} pulses} \label{sec:unipolar}

In 2003, Falcke \& Gorham \citep{FalckeGorham2003} motivated that the radio emission from extensive air showers could be described as ``geosynchrotron radiation'' arising from the acceleration of air shower electrons and positrons in the earth's magnetic field. This approach was afterwards followed by a number of authors with implementations in both the frequency-domain (see section \ref{sec:bipolar}) and time-domain. In principle, calculations in both the frequency-domain and time-domain can be considered equivalent. As the radio signal, however, is very localized in time, a time-domain calculation has a number of technical advantages. Also, the treatment of retardation effects is simpler in the time-domain than in the frequency domain.

The first time-domain implementation of the geosynchrotron approach, for the emission from the air shower maximum only, was made by Suprun et al.\ \citep{SuprunGorhamRosner2003}. In contrast to the frequency-domain calculation carried out in parallel by Huege \& Falcke \citep{HuegeFalcke2003a}, the frequency spectra predicted by the Suprun et al.\ calculation levelled off at very low frequencies and the pulses were \emph{unipolar} (see Figs.\ 4 and 5 in \citep{SuprunGorhamRosner2003}).

\begin{figure*}[htb]
\centering
\includegraphics[angle=270,width=0.49\textwidth]{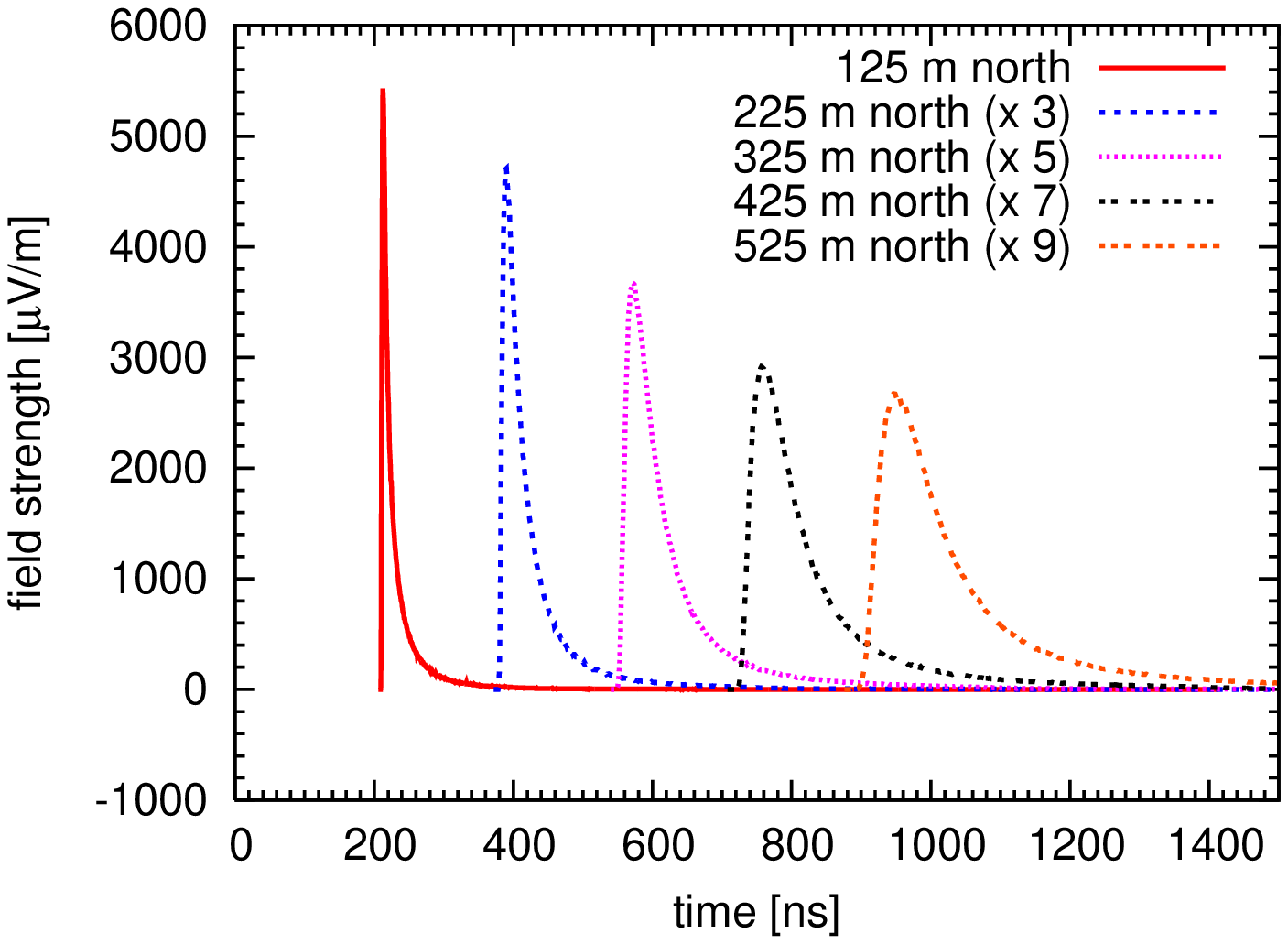}
\includegraphics[angle=270,width=0.49\textwidth]{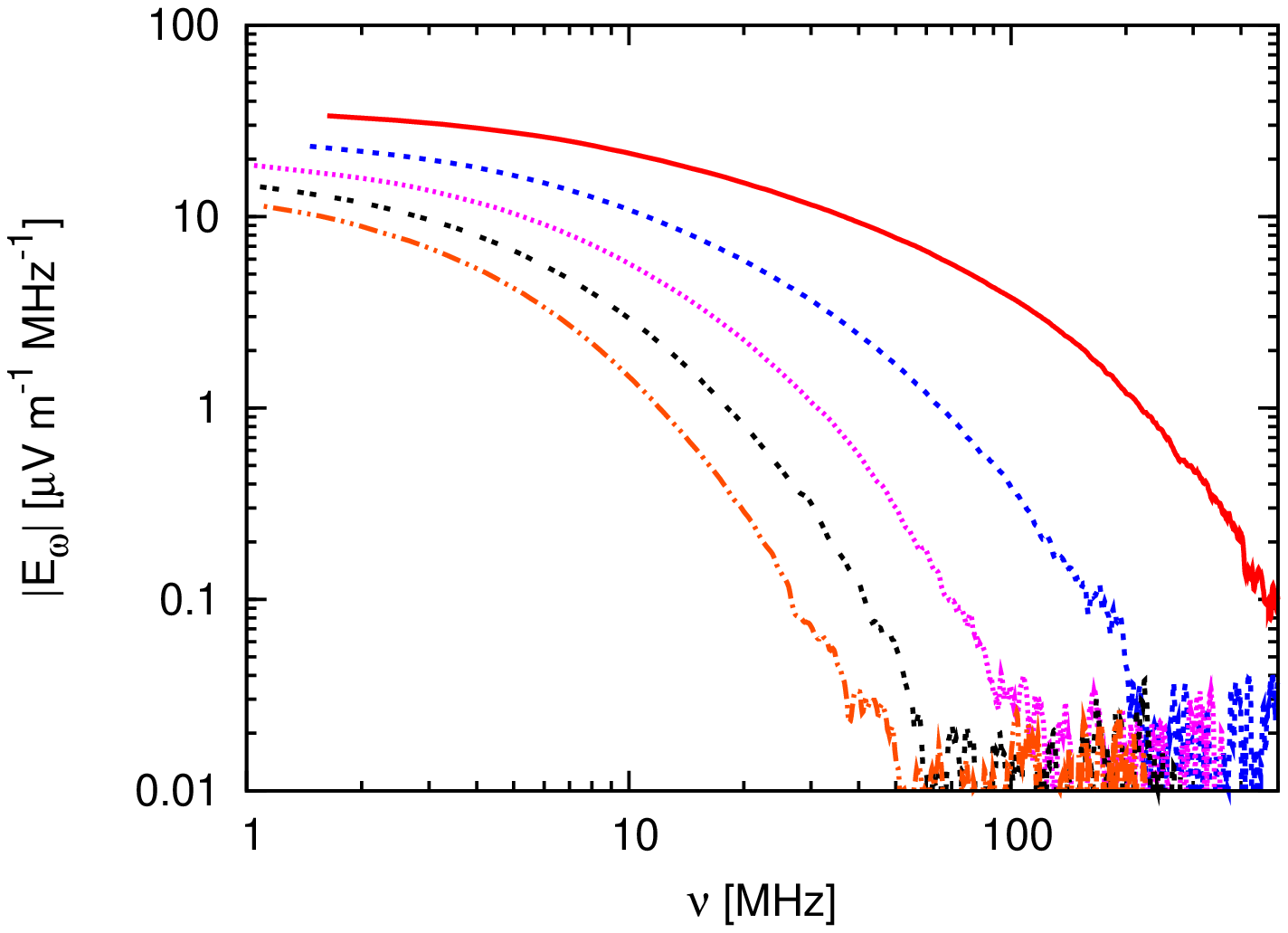}
\caption{\label{fig:reas2}\emph{Unipolar} radio pulses (left) of a $10^{18}$~eV air shower with $30^{\circ}$ zenith angle simulated with REAS2 \citep{HuegeUlrichEngel2007a} and the corresponding frequency spectra (right).}
\end{figure*}

In the following years, much effort was spent on developing a detailed time-domain implementation of the geosynchrotron model by Huege \& Falcke and later Huege, Ulrich \& Engel in a series of publications \citep{HuegeFalcke2005a,HuegeFalcke2005b,HuegeUlrichEngel2007a}. The first implementation (REAS1, \citep{HuegeFalcke2005a}) was based on a parameterized air shower model. Later, a transition was made to realistic air shower characteristics derived on a per-shower basis from CORSIKA \citep{HeckKnappCapdevielle1998} simulations (REAS2, \citep{HuegeUlrichEngel2007a}). A common feature of all of these time-domain calculations was the prediction of \emph{unipolar} pulses with frequency spectra leveling off at very low frequencies, as seen in Fig.\ \ref{fig:reas2} for the case of REAS2.

The geosynchrotron approach was also implemented in the AIRES air shower radio emission code \citep{Sciutto1999} by DuVernois et al.\ \citep{DuVernoisIcrc2005}. This implementation exhibited the same characteristic \emph{unipolar} pulses and frequency spectra leveling off at very low frequencies. However, the pulses predicted by this code were always a factor of 10--20 higher in electric field amplitude than those of REAS2, in spite of the near-identical implementation of the radiation physics.

\begin{figure}[htb]
\centering
\includegraphics[width=0.42\textwidth]{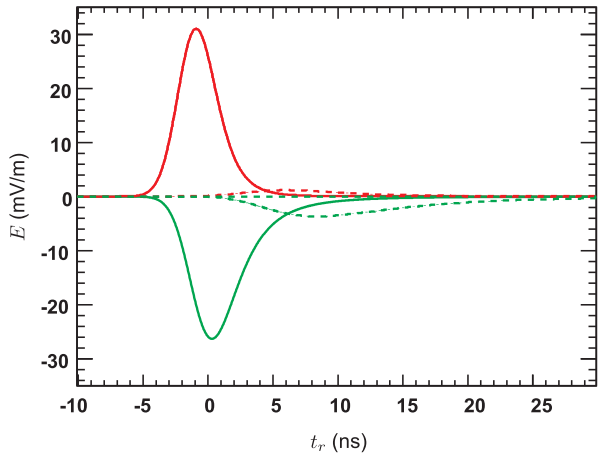}
\caption{\label{fig:ATM}\emph{Unipolar} pulses calculated with the ATM model \citep{ChauvinRiviereMontanet2010}.}
\end{figure}

A recent modelling effort by Chauvin et al. \citep{ChauvinRiviereMontanet2010} is based on an analytic approximation of the radio emission from air showers, described in the time-domain. Its results look strikingly similar to those of the various geosynchrotron implementations. In particular, the model also predicts \emph{unipolar} pulses (see Fig.\ \ref{fig:ATM}) and thus frequency spectra leveling off at very low frequencies. Amplitude-wise, its predictions seem to agree with the AIRES-based implementation of the geosynchrotron model and are thus again a factor of 10--20 higher than those predicted by REAS2 \citep{DuVernoisIcrc2005}.

\subsection{Models with \emph{bipolar} pulses} \label{sec:bipolar}

Motivated by the initial article of Falcke \& Gorham, a frequency-domain calculation of geosynchrotron radiation was carried out by Huege \& Falcke in \citep{HuegeFalcke2003a}. As the calculation was based on synchrotron spectra of individual particles, the resulting integrated frequency spectra showed a fall-off to zero at very low frequencies (cf.\ Figs.\ 6 and 9 in \citep{HuegeFalcke2003a}). The frequency-domain implementation of the geosynchrotron model therefore falls in the group of models with \emph{bipolar} pulses. Due to its analytic nature, however, the model had to employ a number of approximations and simplifications. Further modelling efforts therefore concentrated on an implementation of the geosynchrotron model in the time-domain Monte Carlo code REAS, cf.\ section \ref{sec:unipolar}.

\begin{figure}[htb]
\centering
\includegraphics[width=0.4\textwidth]{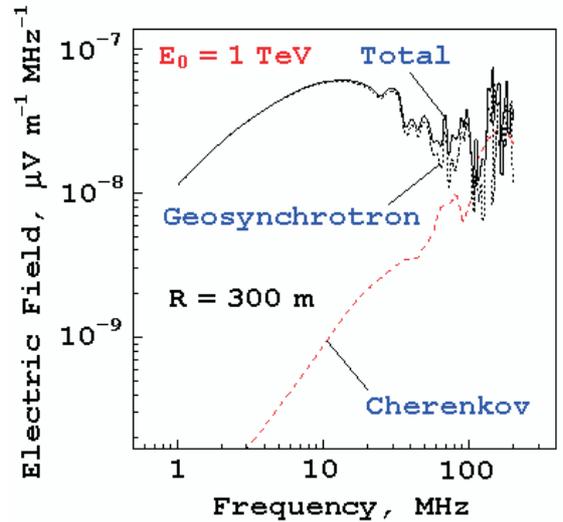}
\caption{\label{fig:kalmykov}Frequency spectra calculated with the EGSnrc-based radio emission code \citep{EngelKalmykovKonstantinovICRC2005}. The spectra drop to zero at very low frequencies, corresponding to \emph{bipolar} pulses in the time-domain.}
\end{figure}

Another approach at modelling the radio emission from air showers was taken by Engel, Kalmykov \& Konstantinov \citep{EngelKalmykovKonstantinovICRC2005}. They implemented a frequency-domain Monte Carlo description in a variant of the EGS code \citep{EGSnrc}, modelling the radiation from straight line-segments and the starts and ends of trajectories in the Fraunhofer limit. The frequency spectra predicted by this approach showed a drop to zero at very low frequencies, visible in Fig.\ \ref{fig:kalmykov}. Again, therefore, a frequency-domain calculation falls into the group of models with \emph{bipolar} pulses.

With their macroscopic geomagnetic radiation model (MGMR, \citep{ScholtenWernerRusydi2008,WernerScholten2008}), Scholten, Werner \& Rusydi developed a macroscopic approach which includes the effects of the induced transverse current, induced dipole moments, charge excess and the influence of a non-unity refractive index. The dominant emission contribution comes from the time-varying transverse currents propagating through the atmosphere with nearly the speed of light. In contrast to the models following the geosynchrotron approach, MGMR does not treat the emission from individual particles, but stresses that coherent emission arises from the motion of charges averaged over appropriately chosen distances in the EAS. Consequently, in a simple approximation where all charges are concentrated in a point \citep{ScholtenWernerRusydi2008}, an analytic formula can be derived to relate the time-structure of the radio pulses directly to the longitudinal evolution of the air shower.

\begin{figure*}[htb]
\centering
\includegraphics[width=0.28\textwidth]{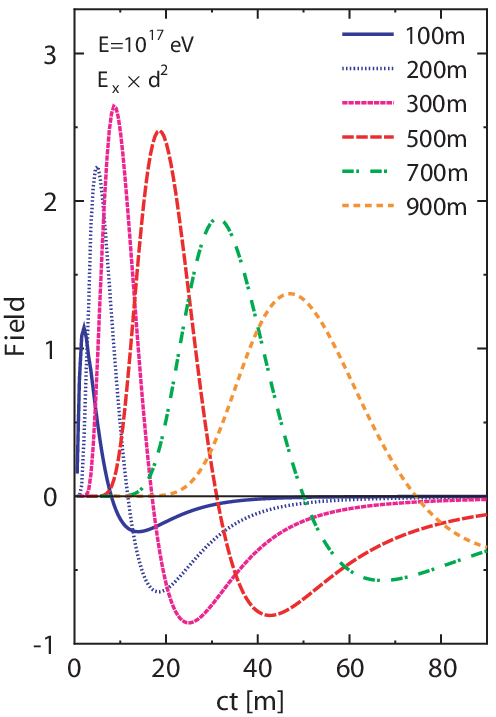}
\includegraphics[width=0.35\textwidth]{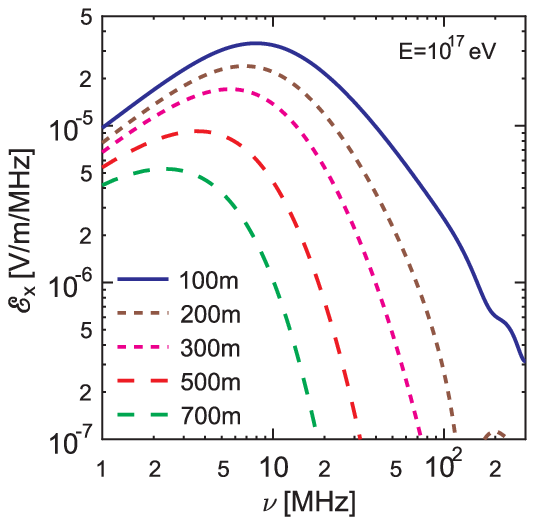}
\caption{\label{fig:mgmr}\emph{Bipolar} pulses (left) and frequency spectra (right) of a $10^{17}$~eV vertical air shower calculated with the MGMR model \citep{ScholtenWernerRusydi2008}.}
\end{figure*}

Although the MGMR model describes the radio emission in the time-domain, it predicts \emph{bipolar} pulses and frequency spectra falling to zero at very low frequencies (see Fig.\ \ref{fig:mgmr}). It was this result that sparked the discussions about discrepancies between models predicting \emph{bipolar} and \emph{unipolar} pulses.

\section{The missing radiation component}

As discussed in the introduction, it became clear from general arguments that the zero-frequency component of the radio emission should vanish \citep{ScholtenWernerARENA2008} and the pulses could therefore not be \emph{unipolar}. Only recently, however, it was identified what caused the discrepancies between the models predicting \emph{unipolar} pulses and those predicting \emph{bipolar} pulses. The reason for the discrepancies is that the models with \emph{unipolar} pulses miss a radiation contribution. All of these models start from the Li\`enard-Wiechert description of the electric field produced by a single moving charged particle,
\begin{eqnarray} \label{eqn:radiate}
\vec{E}(\vec{x},t) &=& e \left[\frac{\vec{n}-\vec{\beta}}{\gamma^{2}(1-\vec{\beta} \cdot \vec{n})^{3} R^2}\right]_{\mathrm{ret}} \nonumber \\
&+& \frac{e}{c} \left[ \frac{\vec{n} \times \{(\vec{n}-\vec{\beta})\times \dot{\vec{\beta}}\}}{(1-\vec{\beta}\cdot\vec{n})^{3}R}\right]_{\mathrm{ret}},
\end{eqnarray}
in which they only take into account the continuous acceleration of particles in the magnetic field.

To calculate the radio emission from an air shower, one has to calculate the radio emission from an ensemble of $N$ relativistic charged particles. The number $N$, however, varies over the air shower evolution and thus has a time-dependence $N(t)$. The models predicting \emph{unipolar} pulses effectively integrate the total radio emission in the air shower as:
\begin{equation}
\vec{E}_{\mathrm{tot}}(\vec{x},t) = N(t) \vec{E}(\vec{x},t)
\end{equation}
This, however, is not correct. Although this equation seems to explicitly take into account the fact that the number of relativistic charged particles is changing with time, a radiation component associated with this variation of the number of charges is neglected. This becomes obvious when recalling that the electric field equation (\ref{eqn:radiate}) itself is derived from the Li\`enard-Wiechert potentials
\begin{eqnarray}
\Phi(\vec{x},t) & = & \left[ \frac{e}{(1 - \vec{\beta} \cdot \vec{n}) R} \right]_{\mathrm{ret}} \nonumber \\
\vec{A}(\vec{x},t) & = & \left[ \frac{e \vec{\beta}}{(1 - \vec{\beta} \cdot \vec{n})} \right]_{\mathrm{ret}} \label{lwpot}
\end{eqnarray}
via
\begin{equation} \label{eqn:fieldderivation}
\vec{E}(\vec{x},t)=-\nabla \Phi(\vec{x},t) - \vec{\dot{A}}(\vec{x},t).
\end{equation}

The fact that not only one charged particle is radiating, but that the number of charged particles $N(t)$ is changing as a function of time has to be taken into account in the calculation of $\vec{A}(\vec{x},t)$ already. The time-derivative applied to calculate $\vec{E}(\vec{x},t)$ then leads to additional radiation terms appearing because of the time-dependence of $N(t)$.

The most straight-forward way to fix the models suffering from this problem is thus to not calculate the electric fields from single particles and integrate over them, but to calculate the integrated vector potential associated with the ensemble of charged particles $\vec{A}_{\mathrm{tot}}(\vec{x},t)$ and only as a last step calculate the electric field $\vec{E}_{\mathrm{tot}}(\vec{x},t)$ via equation (\ref{eqn:fieldderivation}). This approach has been chosen in the MGMR model \citep{ScholtenWernerRusydi2008}. Another option to take into account the missing radiation component has been chosen in REAS3 \citep{LudwigHuege2010,LudwigHuegeARENA2010} and will be discussed in the following section.


\section{Endpoint treatment in REAS3}

\begin{figure}[htb]
\centering
\includegraphics[width=0.25\textwidth]{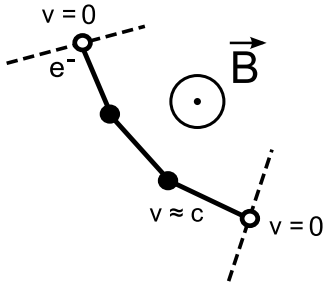}
\caption{\label{fig:REAS3}In REAS3 \citep{LudwigHuege2010}, radio emission is calculated from the acceleration occurring in ``kinks'' joined by straight line segments. At the start and the end of a particle trajectory, additional radiation is generated due to the ``instantaneous'' acceleration of particles from rest to $v\approx c$ and vice-versa.}
\end{figure}

The radio emission component missing in REAS1/2 can be incorporated by considering the radiation associated with the ``endpoints'' of the individual particle tracks in the Monte Carlo summation. This has been realized in REAS3 \citep{LudwigHuege2010,LudwigHuegeARENA2010}. In this implementation, the radio emission from charged particles being deflected in the earth's magnetic field is described as radiation arising from ``kinks'' joined by straight line segments. Most importantly, however, there is additional emission at the starts and ends of the particle tracks associated with the ``instantaneous'' acceleration of particles from rest to their velocity near the speed of light, and the deceleration of particles from their velocity to rest, respectively (cf.\ Fig.\ \ref{fig:REAS3}). If the number of charged particles at a given atmospheric depth changes, there will be a different number of ``start-points'' than ``end-points'', leading to a net contribution caused by the varying number of charges. This effectively takes into account the emission contribution missing in earlier implementations, and leads to \emph{bipolar} pulses and frequency spectra falling to zero at very low frequencies (cf.\ Figs\ \ref{pulsesvertical1e17} and \ref{spectravertical1e17}).

A very similar approach has been chosen in a recent Monte Carlo code for the calculation of radio emission from showers in dense media by Alvarez-Mu\~niz, Romero-Wolf \& Zas \citep{AlvarezMunizRomeroWolfZas2010}. In general, this ``endpoint'' approach has been found to be a powerful way of describing radiation from arbitrary ensembles of moving particles, as has been detailed in an article by James et al.\ \citep{JamesFalckeHuege2010}. The particular strength of this approach is its universality, which does not require any differentiation between emission mechanisms such as synchrotron radiation or charge excess emission. The revised implementation in REAS3 thus incorporates not only synchrotron emission, but describes the complete radio emission arising from the underlying charged particle motion.

\section{A comparison between REAS3 and MGMR}

Now that the discrepancies between the geosynchrotron approach and the macroscopic geomagnetic radiation model have been resolved, we proceed to make a detailed comparison between REAS3 and MGMR radio emission simulations.

\subsection{Methodology}

To carry out a direct comparison, we have defined a set of prototype air showers with a fixed geometry, a specific particle energy and a fixed set of observer locations. Shower-to-shower fluctuations have been removed by choosing one typical air shower from a set of CORSIKA simulations which was consecutively used as the basis for both the REAS3 and MGMR simulations. The magnetic field and observer altitude were configured with values appropriate to the southern site of the Pierre Auger Observatory. It should be noted that the REAS3 simulations shown here exhibit a slightly elevated level of numerical noise. This applies mostly to near-vertical air showers and will be improved in the release version of the code. In the MGMR model, numerical noise is less of an issue since the motion of charges is averaged in the initial stage of the calculation of coherent radiation, and afterwards the field is calculated.

\subsection{Vertical $10^{17}$ eV showers}

\begin{figure*}[htb]
\begin{minipage}{0.49\textwidth}
\centering
\includegraphics[angle=270,width=\textwidth]{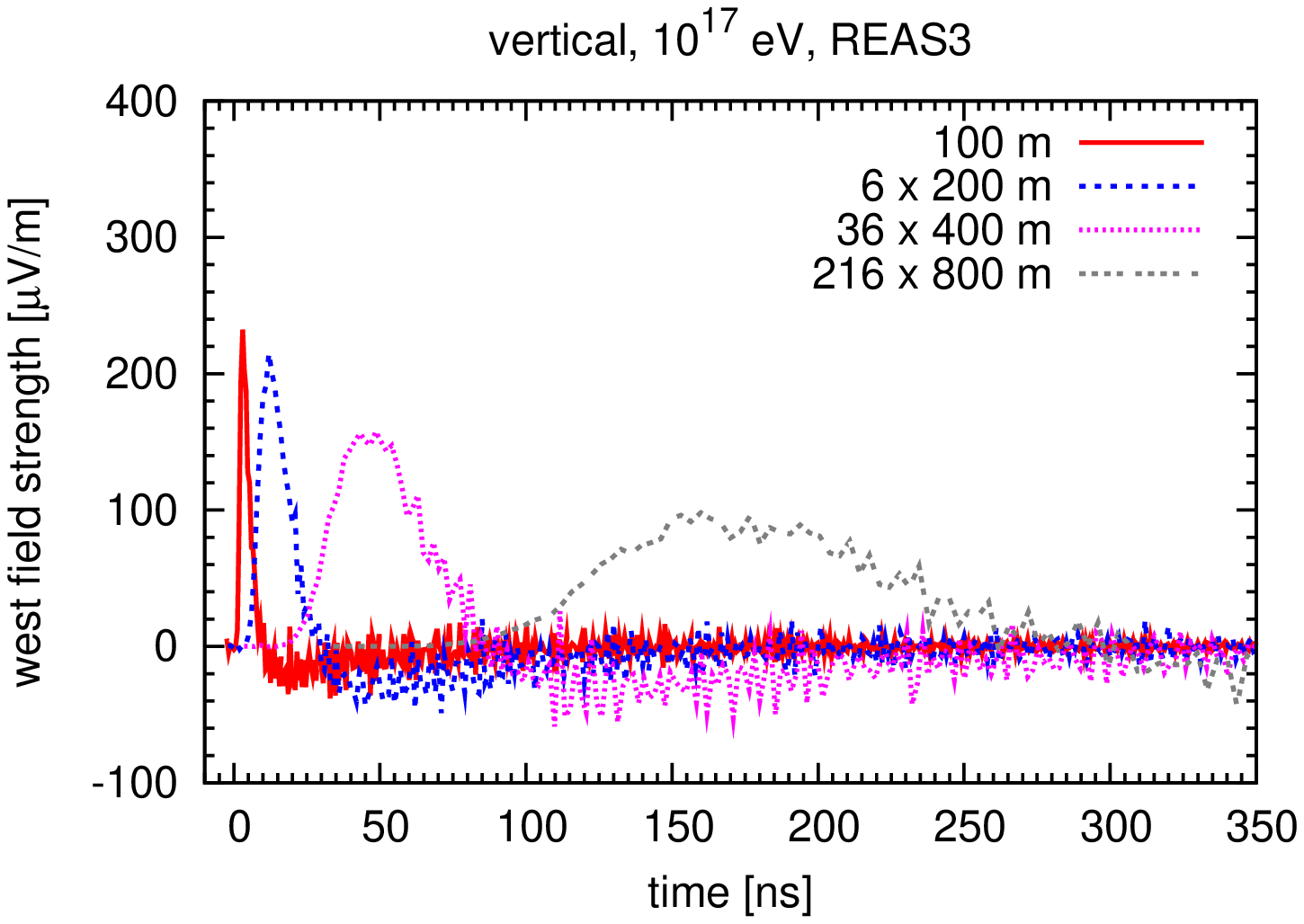}
\end{minipage} \hspace{1.5pc}
\begin{minipage}{0.49\textwidth}
\includegraphics[angle=270,width=\textwidth]{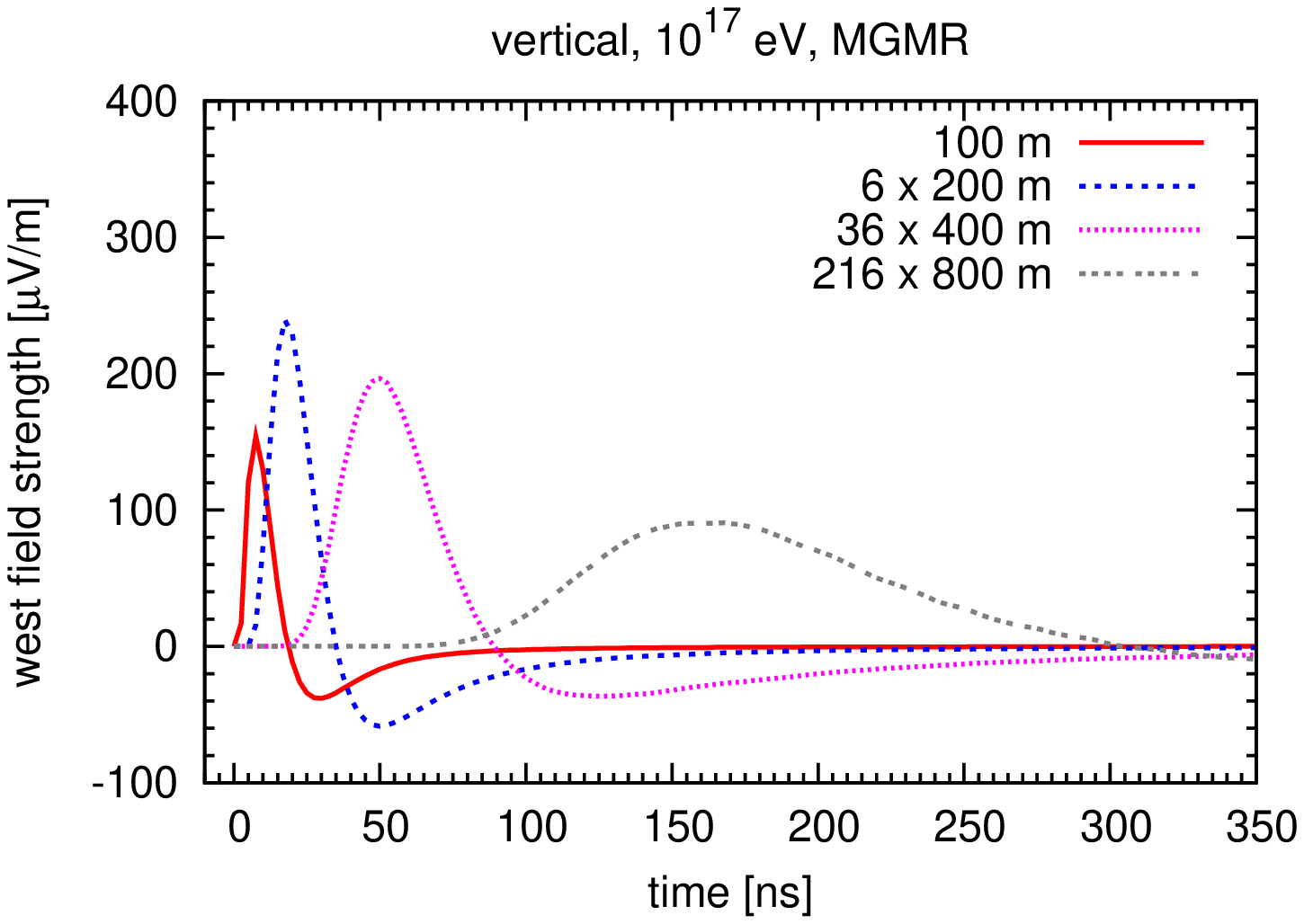}
\end{minipage}
\caption{\label{pulsesvertical1e17}Comparison of the western polarization unlimited bandwidth radio pulses for vertical showers with a primary energy of $10^{17}$ eV for REAS3 (left) and MGMR (right).}
\end{figure*}

\begin{figure}[htb]
\centering
\includegraphics[angle=270,width=0.49\textwidth]{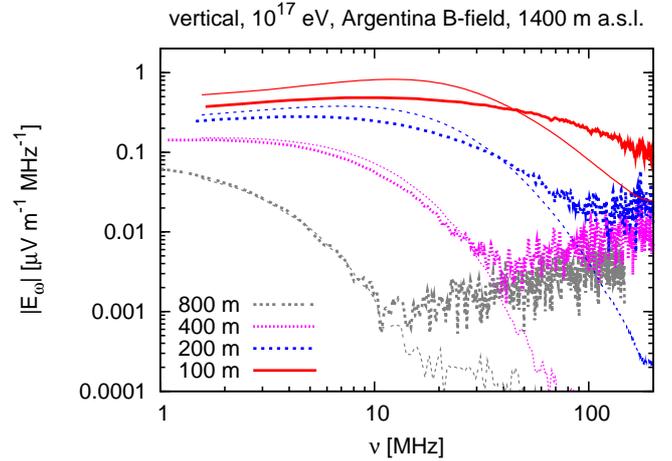}
\caption{\label{spectravertical1e17}Comparison of the absolute value frequency spectra for vertical showers with a primary energy of $10^{17}$ eV for REAS3 (thick lines) and MGMR (thin lines).}
\end{figure}

\begin{figure*}[htb]
\centering
\includegraphics[angle=270,width=0.31\textwidth]{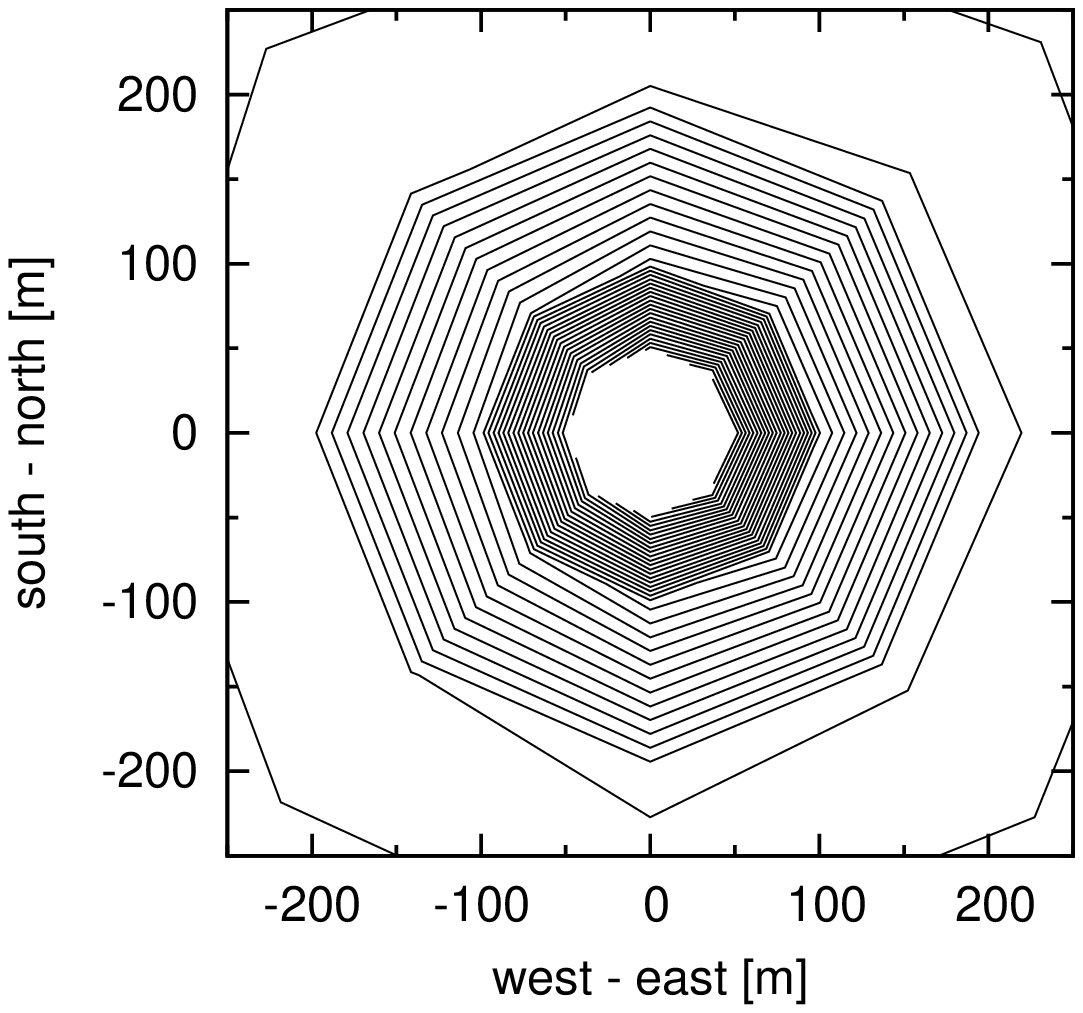}
\includegraphics[angle=270,width=0.31\textwidth]{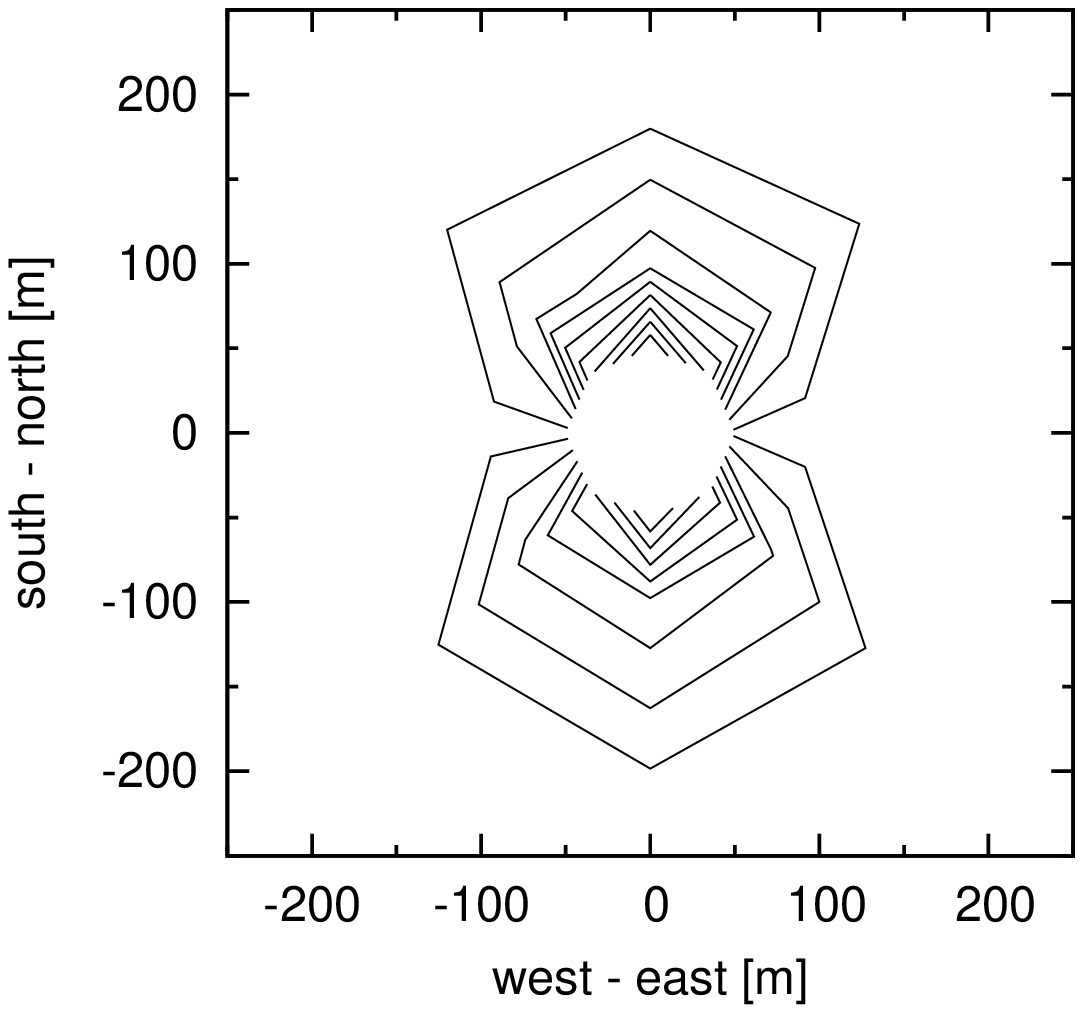}
\includegraphics[angle=270,width=0.31\textwidth]{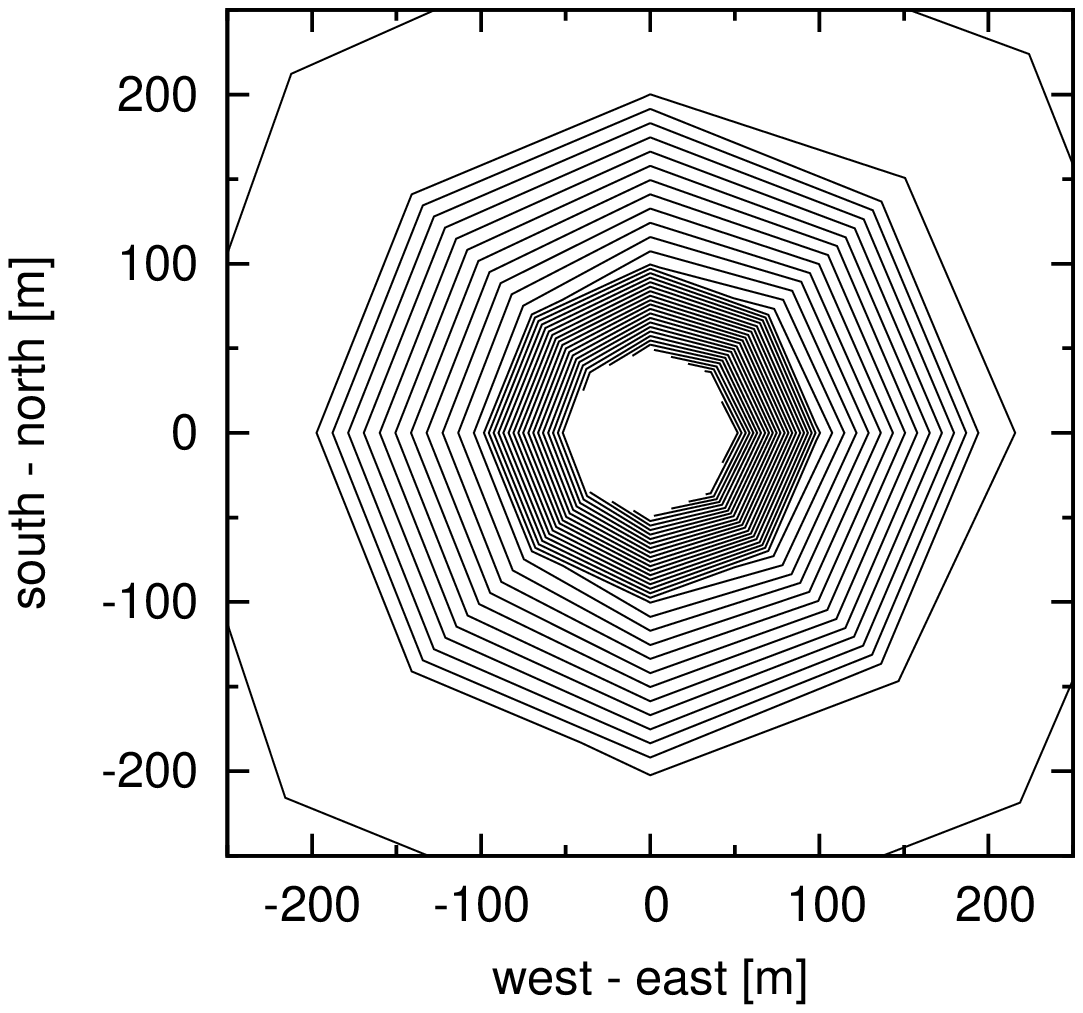}
\\
\includegraphics[angle=270,width=0.31\textwidth]{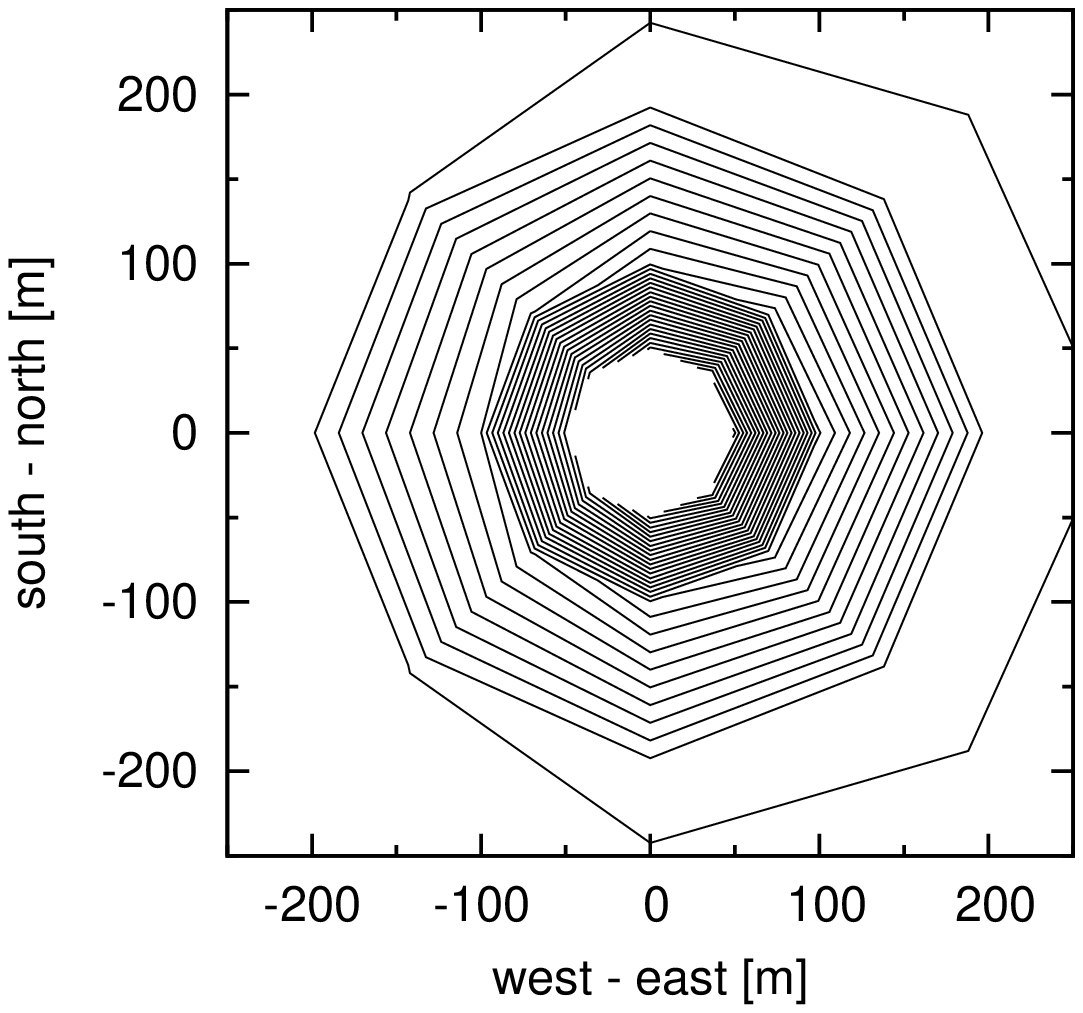}
\includegraphics[angle=270,width=0.31\textwidth]{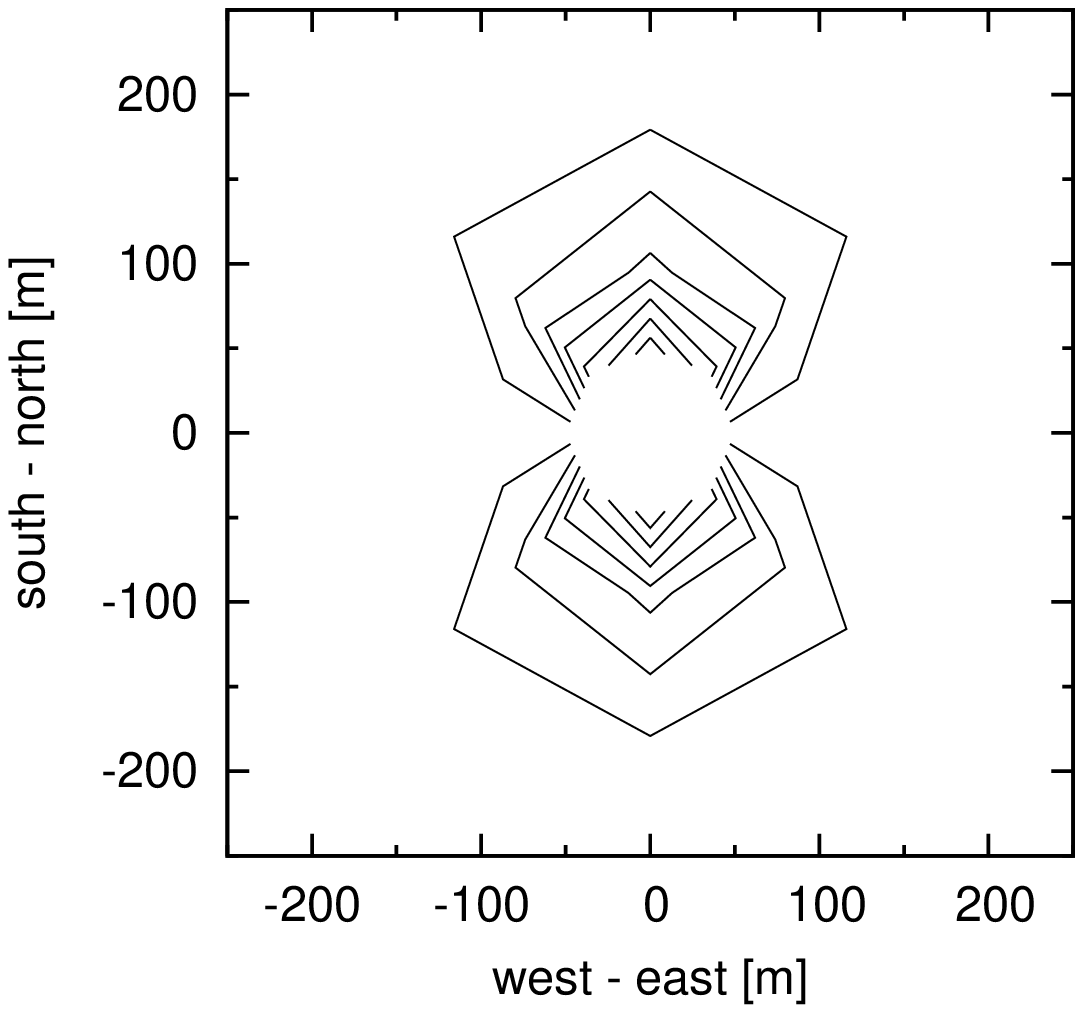}
\includegraphics[angle=270,width=0.31\textwidth]{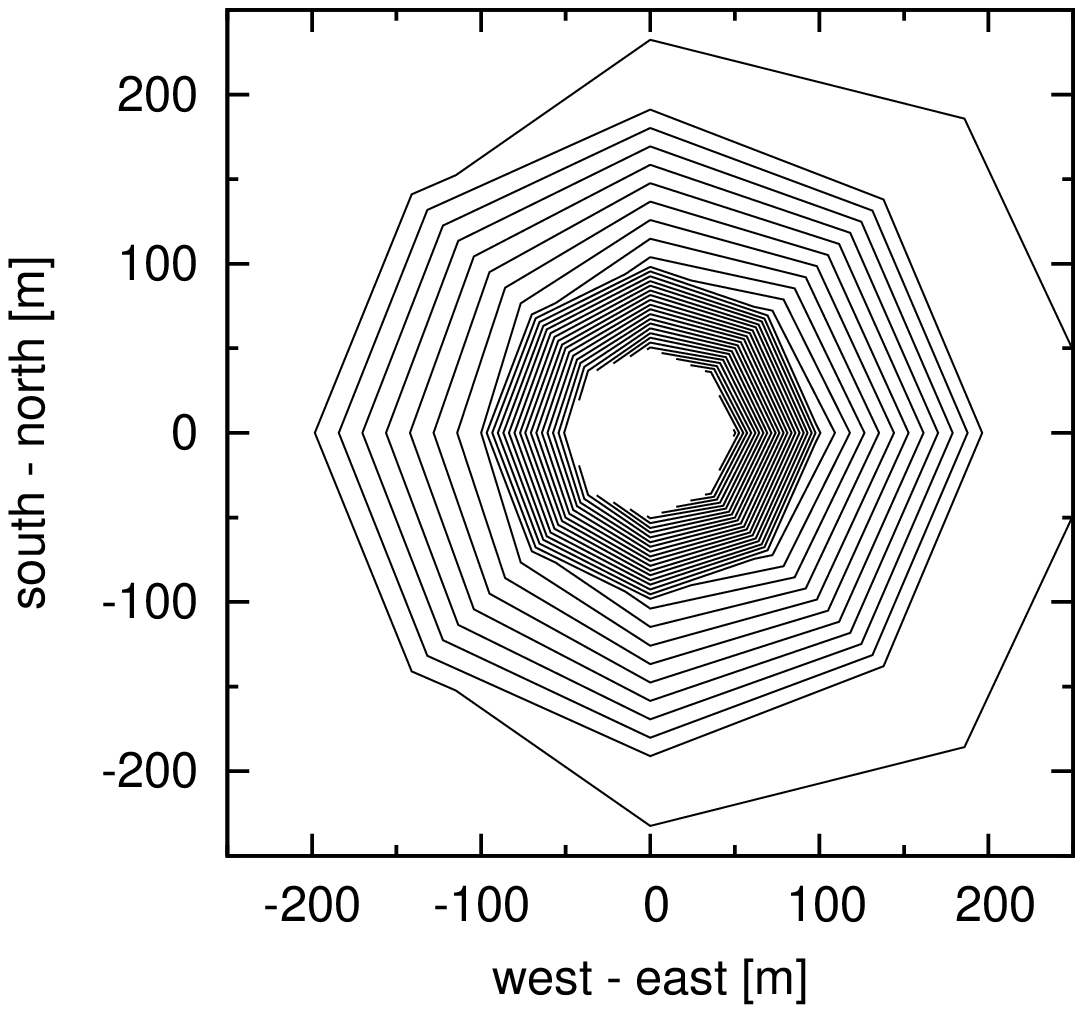}
\caption{\label{contoursvertical1e17}Contour plots of the 60~MHz spectral electric field for vertical showers with a primary energy of $10^{17}$ eV for REAS3 (top) and MGMR (bottom). From left to right: absolute field strength, north polarization component, west polarization component. The white region in the center has not been simulated.}
\end{figure*}

We first compare radio pulses emitted by a vertical $10^{17}$~eV air shower in Fig.\ \ref{pulsesvertical1e17}. Both the \emph{bipolar} pulse shapes and the amplitudes agree remarkably well, with slight deviations close to the shower axis. Likewise, the agreement between the frequency spectra depicted in Fig.\ \ref{spectravertical1e17} is very good with only slight discrepancies at small distances from the shower axis.
\begin{figure*}[htb]
\begin{minipage}{0.49\textwidth}
\centering
\includegraphics[angle=270,width=\textwidth]{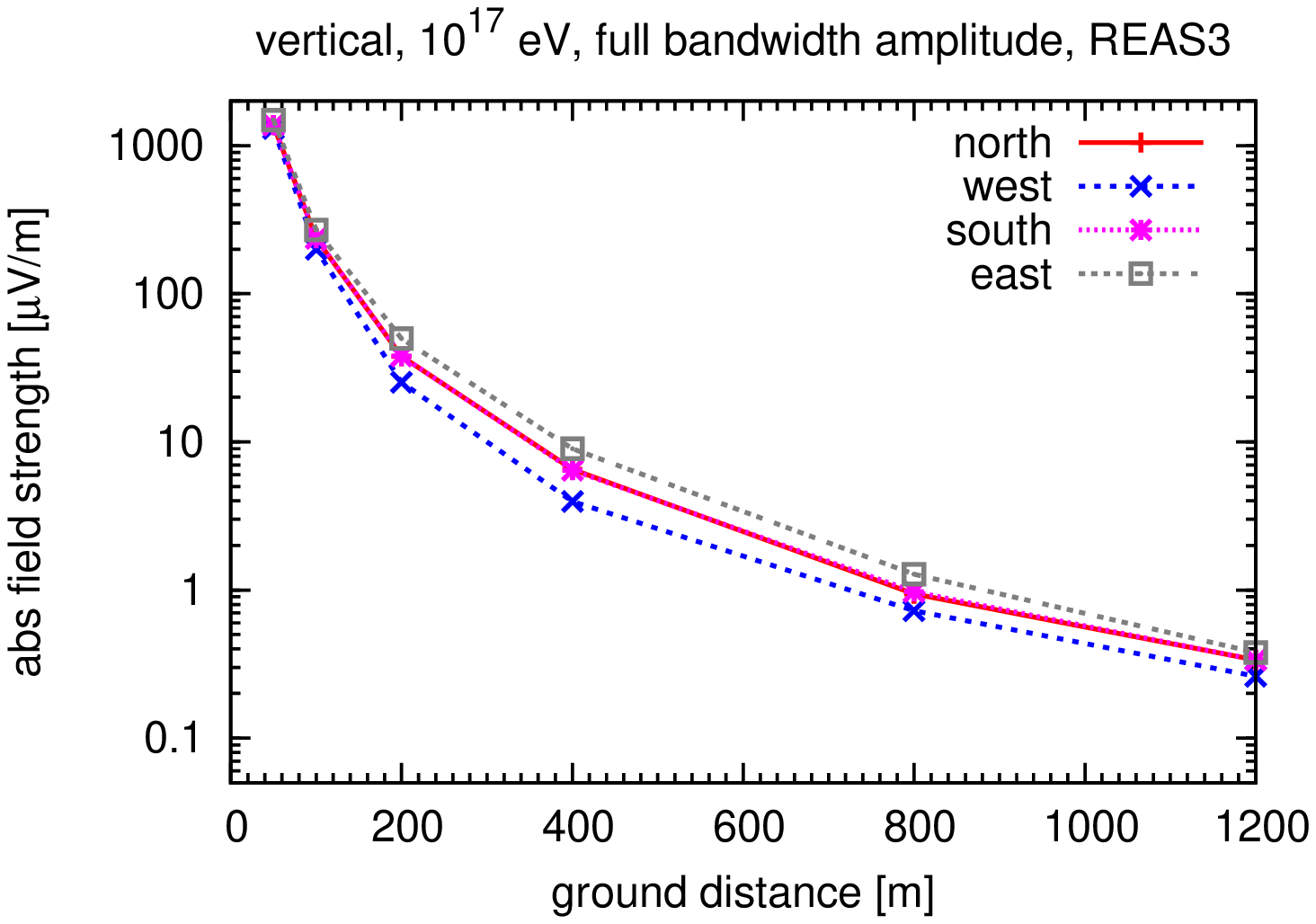}
\end{minipage} \hspace{1.5pc}
\begin{minipage}{0.49\textwidth}
\includegraphics[angle=270,width=\textwidth]{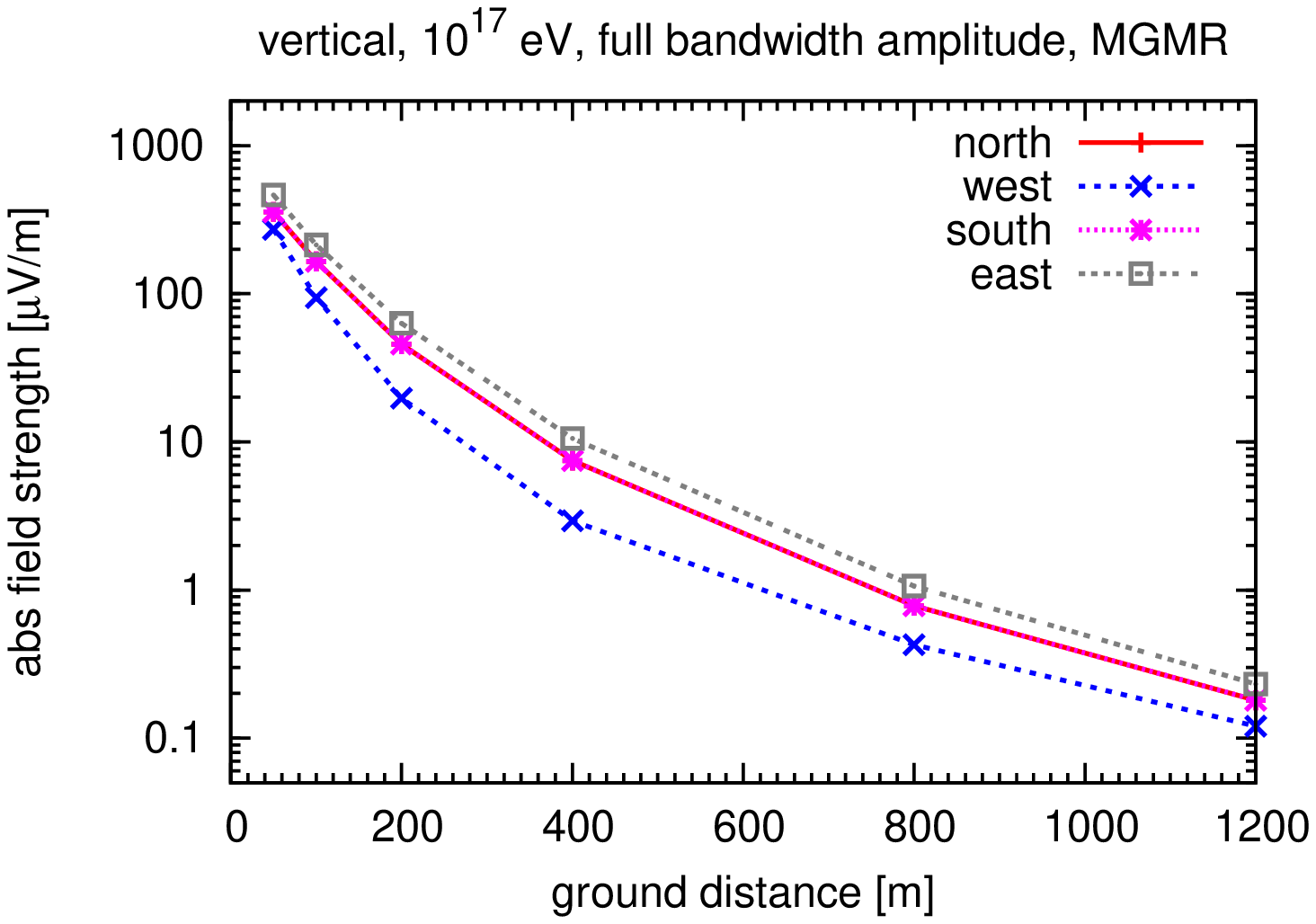}
\end{minipage}
\caption{\label{lateralvertical1e17}Comparison of the absolute field strength unlimited bandwidth pulse amplitude lateral distribution for vertical showers with a primary energy of $10^{17}$ eV for REAS3 (left) and MGMR (right).}
\end{figure*}
A look at the lateral distributions of the unfiltered (i.e., unlimited bandwidth) radio pulse amplitudes in Fig.\ \ref{lateralvertical1e17} shows that both models exhibit a noticeable east-west asymmetry. This is caused by the variation of the net charge excess in the air shower, causing a radiation contribution with a radial polarization signature \citep{LudwigHuege2010,LudwigHuegeARENA2010,DeVriesARENA2010,DeVriesScholtenWerner2010}. It is thus clear that the radio emission from extensive air showers cannot follow a pure geomagnetic $\vec{v} \times \vec{B}$ polarization. The same signature can be seen in contour plots of the 60~MHz spectral emission component shown in Fig.\ \ref{contoursvertical1e17}, which would exhibit a strict east-west polarization for pure geomagnetic emission. The presence of such a deviation from purely geomagnetic emission is currently being investigated with experimental data, as presented for example in \citep{SchoorlemmerARENA2010}.

\subsection{No magnetic field}

\begin{figure*}[h!tb]
\begin{minipage}{0.49\textwidth}
\centering
\includegraphics[angle=270,width=\textwidth]{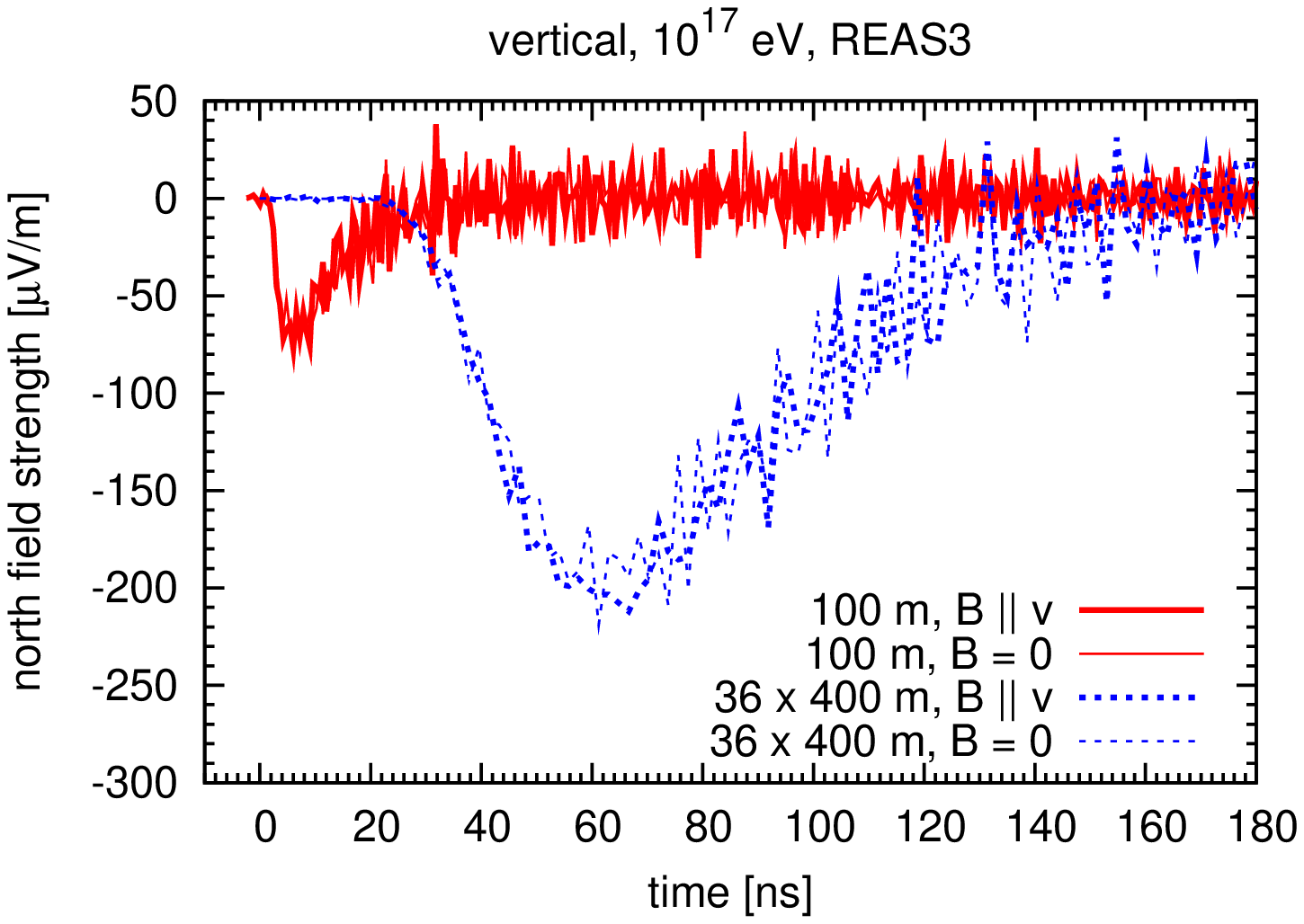}
\end{minipage} \hspace{1.5pc}
\begin{minipage}{0.49\textwidth}
\includegraphics[angle=270,width=\textwidth]{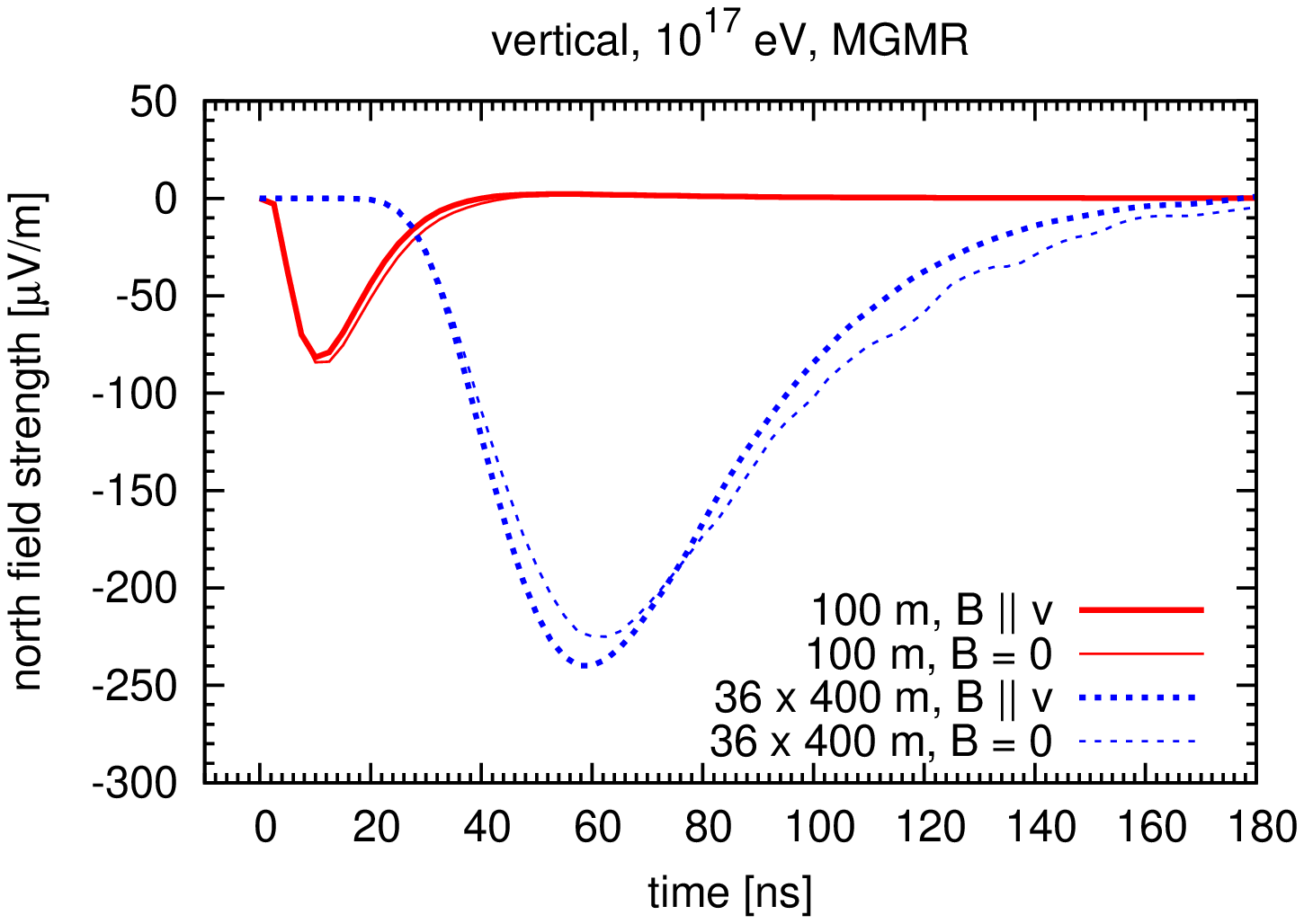}
\end{minipage}
\caption{\label{pulsesnobfieldvertical1e17}Pulses for vertical $10^{17}$ eV in absence of magnetic field or magnetic field parallel to the shower axis for REAS3 (left) and MGMR (right).}
\end{figure*}

If the magnetic field is switched off completely or is chosen parallel to the air shower axis in both REAS3 and MGMR, only the pure radio emission from the varying charge excess remains. It shows the expected radial polarization characteristics and is extremely similar for both models as shown in Fig.\ \ref{pulsesnobfieldvertical1e17}. This once more clearly illustrates that the ``endpoint'' approach implemented in REAS3 universally predicts the complete radio emission, even the components not associated with magnetic field effects, in agreement with the MGMR predictions.
%

\subsection{Inclined showers}

\begin{figure*}[h!tb]
\begin{minipage}{0.49\textwidth}
\centering
\includegraphics[angle=270,width=\textwidth]{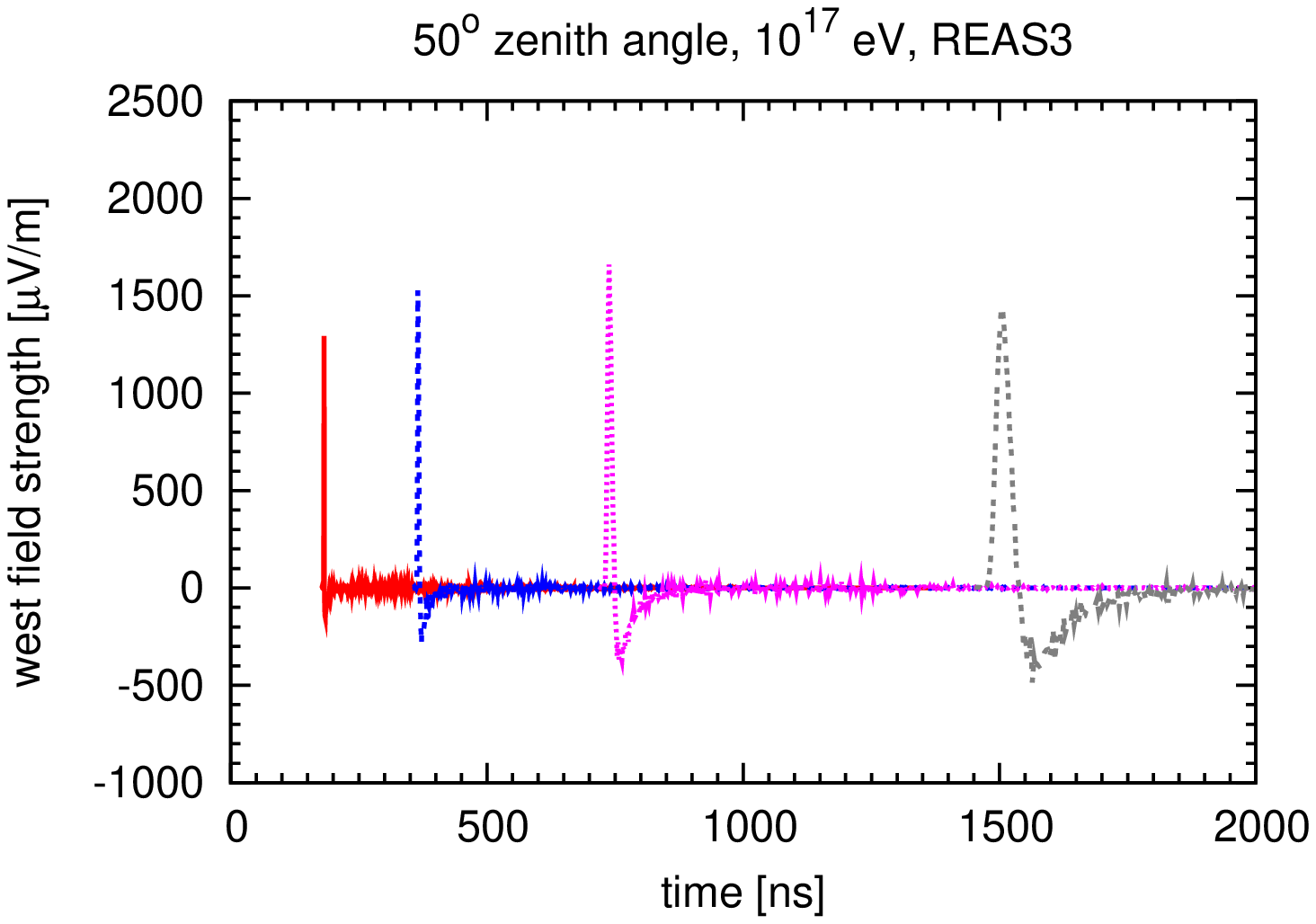}
\end{minipage} \hspace{1.5pc}
\begin{minipage}{0.49\textwidth}
\includegraphics[angle=270,width=\textwidth]{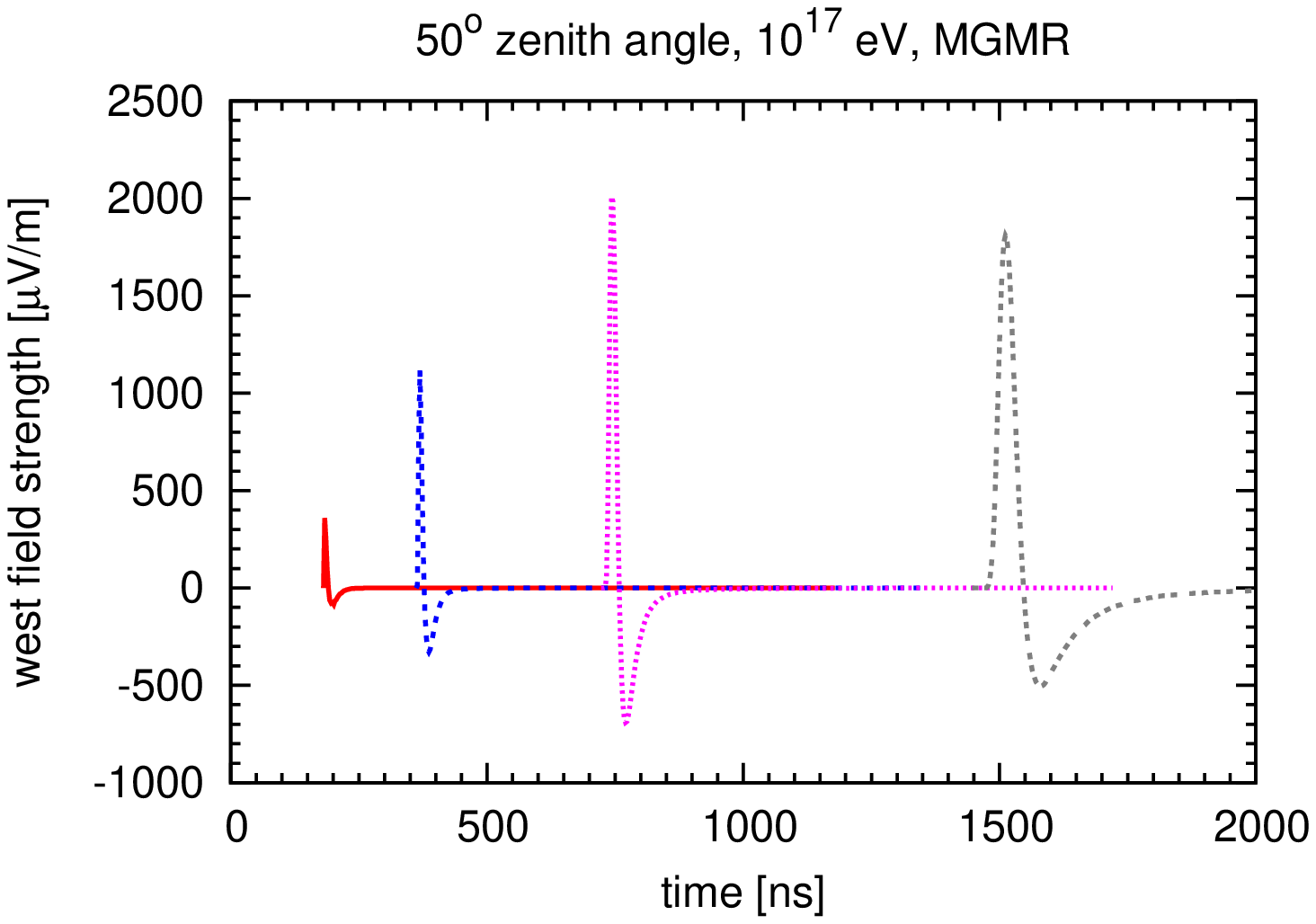}
\end{minipage}
\caption{\label{pulsesinclined1e17}Pulses of the western polarization component for a $10^{17}$ eV shower with 50$^{\circ}$ zenith angle for REAS3 (left) and MGMR (right). The pulses (from left to right) denote the field strengths at 100~m, 6 times the field strengths at 200~m, 36 times the field strengths at 400~m and 216 times the field strengths at 800~m.}
\end{figure*}

In Fig.\ \ref{pulsesinclined1e17}, radio pulses from a $10^{17}$~eV air shower with 50$^{\circ}$ zenith angle are compared between REAS3 and MGMR. A noticeable discrepancy between the two models arises close to the shower core, where REAS3 predicts significantly larger amplitudes than MGMR. This is not surprising, as close to the shower axis, details of the air shower model become increasingly important. The MGMR model currently uses a somewhat simplified parameterization for the thickness of the air shower ``pancake'' and neglects the lateral distribution of particles. For inclined air showers, this has a stronger influence than for near-vertical showers, as identical ground distances correspond to smaller effective axis distances.
\begin{figure*}[htb]
\begin{minipage}{0.49\textwidth}
\centering
\includegraphics[angle=270,width=\textwidth]{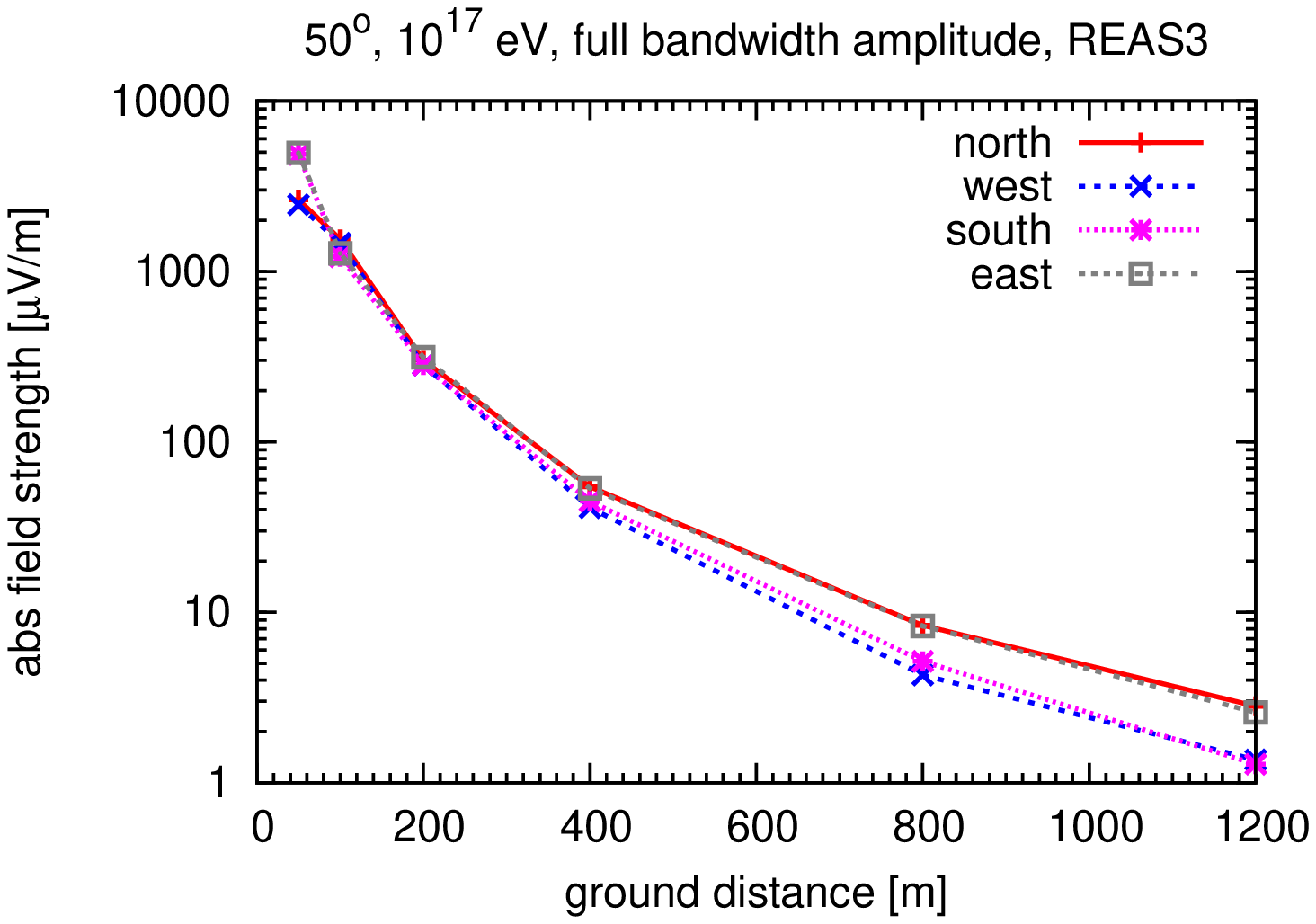}
\end{minipage} \hspace{1.5pc}
\begin{minipage}{0.49\textwidth}
\includegraphics[angle=270,width=\textwidth]{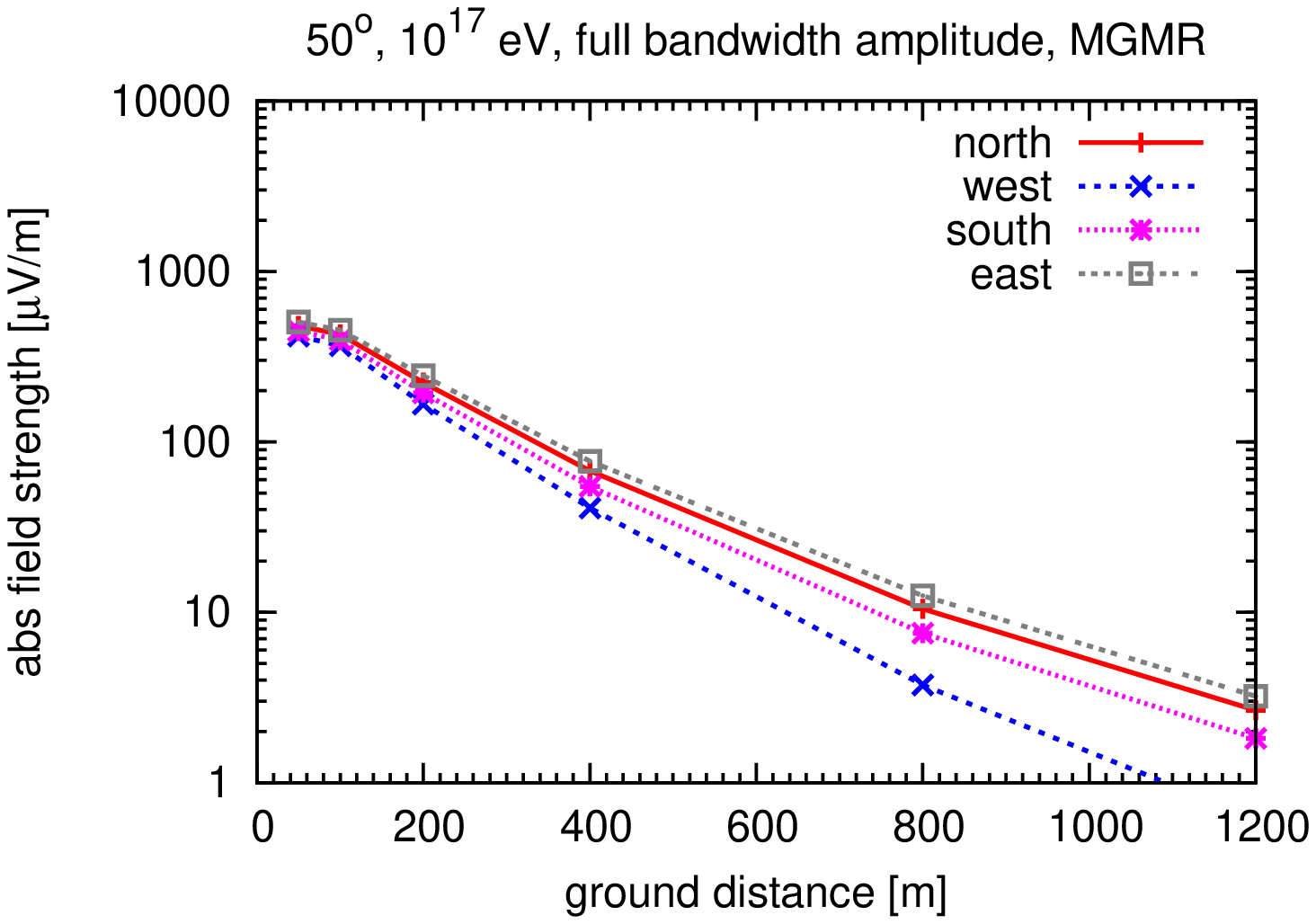}
\end{minipage}
\caption{\label{lateralinclined1e17}Lateral distribution of the unlimited bandwidth absolute pulse amplitude for a $10^{17}$ eV shower with 50$^{\circ}$ zenith angle for REAS3 (left) and MGMR (right).}
\end{figure*}
The deviations of the two models near the shower core can also be seen in the unfiltered (unlimited bandwidth) lateral distributions of the radio pulse amplitudes shown in Fig.\ \ref{lateralinclined1e17}.






\section{Conclusions}

After many years of research, the modelling of radio emission from cosmic ray air showers has achieved a true breakthrough. Discrepancies between models predicting \emph{unipolar} pulses and models predicting \emph{bipolar} pulses have been reconciled. They were caused by a missing radio emission contribution arising from the variation of the number of charged particles during the air shower evolution. This emission component can be incorporated in existing models by calculating the vector potential and deriving the electric field only as a last step in the calculation, as has always been the case in the MGMR model. Alternatively, a powerful and universal approach using ``endpoints'' can be used to take the missing radiation into account. The latter approach has been followed in REAS3. Comparisons between REAS3 and the MGMR model confirm that the agreement between these two completely independent models based on very different approaches is indeed good. This is a major milestone, since previously different models predicted both qualitatively and quantitatively vastly different results. Remaining discrepancies in the radio emission amplitudes close to the shower core predicted by REAS3 and MGMR can be attributed to differences in the underlying air shower model, which is parameterized in MGMR while it is coming from CORSIKA-derived histograms in case of REAS3.

\section*{Acknowledgements}

The authors would like to thank many colleagues for useful discussions, in particular S.\ Buitink, H.\ Falcke, C.W.\ James and Klaus Werner. This research has been supported by grant number VH-NG-413 of the Helmholtz Association.

\bibliographystyle{model1-num-names}

\begin{thebibliography}{25}
\expandafter\ifx\csname natexlab\endcsname\relax\def\natexlab#1{#1}\fi
\providecommand{\bibinfo}[2]{#2}
\ifx\xfnm\relax \def\xfnm[#1]{\unskip,\space#1}\fi
\bibitem[{{Huege}(2009)}]{HuegeArena2008}
\bibinfo{author}{T.~{Huege}},
\newblock \bibinfo{title}{{Simulations and theory of radio emission from cosmic
  ray air showers}},
\newblock \bibinfo{journal}{Nucl. Instr. Meth. A} \bibinfo{volume}{604}
  (\bibinfo{year}{2009}) \bibinfo{pages}{S57--S63}.
\bibitem[{Rad(2007)}]{RadioTheoryMeeting2007}
  \bibinfo{note}{http://www-ik.fzk.de/\~{}huege/theorymeeting.html}.
\bibitem[{{Gousset} et~al.(2009){Gousset}, {Lamblin}, and
  {Valcares}}]{GoussetLamblinValcares2009}
\bibinfo{author}{T.~{Gousset}}, \bibinfo{author}{J.~{Lamblin}},
  \bibinfo{author}{S.~{Valcares}},
\newblock \bibinfo{title}{{Radioelectric fields from cosmic-ray air showers at
  large impact parameters}},
\newblock \bibinfo{journal}{Astropart. Phys.} \bibinfo{volume}{31}
  (\bibinfo{year}{2009}) \bibinfo{pages}{52--62}.
\bibitem[{{Scholten} and {Werner}(2009)}]{ScholtenWernerARENA2008}
\bibinfo{author}{O.~{Scholten}}, \bibinfo{author}{K.~{Werner}},
\newblock \bibinfo{title}{{Macroscopic model of geomagnetic-radiation from air
  showers}},
\newblock \bibinfo{journal}{Nucl. Instr. Meth. A} \bibinfo{volume}{604}
  (\bibinfo{year}{2009}) \bibinfo{pages}{S24--S26}.
\bibitem[{{Ludwig} and {Huege}(2010{\natexlab{a}})}]{LudwigHuege2010}
\bibinfo{author}{M.~{Ludwig}}, \bibinfo{author}{T.~{Huege}},
\newblock \bibinfo{title}{{REAS3: A revised implementation of the
  geosynchrotron model for radio emission from EAS}},
\newblock \bibinfo{journal}{Astropart. Phys.} \bibinfo{volume}{submitted}
  (\bibinfo{year}{2010}{\natexlab{a}}).
\bibitem[{{Ludwig} and {Huege}(2010{\natexlab{b}})}]{LudwigHuegeARENA2010}
\bibinfo{author}{M.~{Ludwig}}, \bibinfo{author}{T.~{Huege}},
\newblock \bibinfo{title}{{REAS3: A revised implementation of the
  geosynchrotron model for radio emission from air showers}},
\newblock \bibinfo{journal}{these proceedings}
  (\bibinfo{year}{2010}{\natexlab{b}}).
\bibitem[{{Scholten} et~al.(2008){Scholten}, {Werner}, and
  {Rusydi}}]{ScholtenWernerRusydi2008}
\bibinfo{author}{O.~{Scholten}}, \bibinfo{author}{K.~{Werner}},
  \bibinfo{author}{F.~{Rusydi}},
\newblock \bibinfo{title}{{A macroscopic description of coherent geo-magnetic
  radiation from cosmic-ray air showers}},
\newblock \bibinfo{journal}{Astroparticle Physics} \bibinfo{volume}{29}
  (\bibinfo{year}{2008}) \bibinfo{pages}{94--103}.
\bibitem[{{de Vries} et~al.(2010){de Vries}, {Scholten}, and
  {Werner}}]{DeVriesARENA2010}
\bibinfo{author}{K.~D. {de Vries}}, \bibinfo{author}{O.~{Scholten}},
  \bibinfo{author}{K.~{Werner}},
\newblock \bibinfo{title}{{Modeling radio signals from Extensive Air Showers}},
\newblock \bibinfo{journal}{these proceedings}  (\bibinfo{year}{2010}).
\bibitem[{{Falcke} and {Gorham}(2003)}]{FalckeGorham2003}
\bibinfo{author}{H.~{Falcke}}, \bibinfo{author}{P.~W. {Gorham}},
\newblock \bibinfo{title}{{Detecting radio emission from cosmic ray air showers
  and neutrinos with a digital radio telescope}},
\newblock \bibinfo{journal}{Astropart. Physics} \bibinfo{volume}{19}
  (\bibinfo{year}{2003}) \bibinfo{pages}{477--494}.
\bibitem[{{Suprun} et~al.(2003){Suprun}, {Gorham}, and
  {Rosner}}]{SuprunGorhamRosner2003}
\bibinfo{author}{D.~A. {Suprun}}, \bibinfo{author}{P.~W. {Gorham}},
  \bibinfo{author}{J.~L. {Rosner}},
\newblock \bibinfo{title}{{Synchrotron radiation at radio frequencies from
  cosmic ray air showers}},
\newblock \bibinfo{journal}{Astropart. Physics} \bibinfo{volume}{20}
  (\bibinfo{year}{2003}) \bibinfo{pages}{157--168}.
\bibitem[{{Huege} and {Falcke}(2003)}]{HuegeFalcke2003a}
\bibinfo{author}{T.~{Huege}}, \bibinfo{author}{H.~{Falcke}},
\newblock \bibinfo{title}{{Radio emission from cosmic ray air showers. Coherent
  geosynchrotron radiation}},
\newblock \bibinfo{journal}{Astronomy \& Astrophysics} \bibinfo{volume}{412}
  (\bibinfo{year}{2003}) \bibinfo{pages}{19--34}.
\bibitem[{{Huege} and {Falcke}(2005{\natexlab{a}})}]{HuegeFalcke2005a}
\bibinfo{author}{T.~{Huege}}, \bibinfo{author}{H.~{Falcke}},
\newblock \bibinfo{title}{{Radio emission from cosmic ray air showers. Monte
  Carlo simulations}},
\newblock \bibinfo{journal}{Astronomy \& Astrophysics} \bibinfo{volume}{430}
  (\bibinfo{year}{2005}{\natexlab{a}}) \bibinfo{pages}{779--798}.
\bibitem[{{Huege} and {Falcke}(2005{\natexlab{b}})}]{HuegeFalcke2005b}
\bibinfo{author}{T.~{Huege}}, \bibinfo{author}{H.~{Falcke}},
\newblock \bibinfo{title}{{Radio emission from cosmic ray air showers:
  Simulation results and parametrization}},
\newblock \bibinfo{journal}{Astropart. Phys.} \bibinfo{volume}{24}
  (\bibinfo{year}{2005}{\natexlab{b}}) \bibinfo{pages}{116}.
\bibitem[{{Huege} et~al.(2007){Huege}, {Ulrich}, and
  {Engel}}]{HuegeUlrichEngel2007a}
\bibinfo{author}{T.~{Huege}}, \bibinfo{author}{R.~{Ulrich}},
  \bibinfo{author}{R.~{Engel}},
\newblock \bibinfo{title}{{Monte Carlo simulations of geosynchrotron radio
  emission from CORSIKA-simulated air showers}},
\newblock \bibinfo{journal}{Astropart. Physics} \bibinfo{volume}{27}
  (\bibinfo{year}{2007}) \bibinfo{pages}{392--405}.
\bibitem[{{Heck} et~al.(1998){Heck}, {Knapp}, {Capdevielle}, {Schatz}, and
  {Thouw}}]{HeckKnappCapdevielle1998}
\bibinfo{author}{D.~{Heck}}, \bibinfo{author}{J.~{Knapp}},
  \bibinfo{author}{J.~N. {Capdevielle}}, \bibinfo{author}{et al.},
  \bibinfo{title}{{CORSIKA: A Monte Carlo Code to
  Simulate Extensive Air Showers}}, \bibinfo{type}{FZKA Report}
  \bibinfo{number}{6019}, Forschungszentrum Karlsruhe, \bibinfo{year}{1998}.
\bibitem[{{Sciutto}(1999)}]{Sciutto1999}
\bibinfo{author}{S.~J. {Sciutto}},
\newblock \bibinfo{title}{{AIRES: A system for air shower simulations (Version
  2.2.0)}},
\newblock \bibinfo{journal}{astro-ph/9911331}  (\bibinfo{year}{1999}).
\bibitem[{{DuVernois} et~al.(2005){DuVernois}, {Cai}, and
  {Kleckner}}]{DuVernoisIcrc2005}
\bibinfo{author}{M.~A. {DuVernois}}, \bibinfo{author}{B.~{Cai}},
  \bibinfo{author}{D.~{Kleckner}},
\newblock \bibinfo{title}{{Geosynchrotron radio pulse emission from extensive
  air showers: Simulations with AIRES}},
\newblock in: \bibinfo{booktitle}{Proc. of the 29th ICRC, Pune, India}, pp.
  \bibinfo{pages}{311--+}.
\bibitem[{{Chauvin} et~al.(2010){Chauvin}, {Rivi{\`e}re}, {Montanet}, {Lebrun},
  and {Revenu}}]{ChauvinRiviereMontanet2010}
\bibinfo{author}{J.~{Chauvin}}, \bibinfo{author}{C.~{Rivi{\`e}re}},
  \bibinfo{author}{F.~{Montanet}}, \bibinfo{author}{et al.},
\newblock \bibinfo{title}{{Radio emission in a toy model with point-charge-like
  air showers}},
\newblock \bibinfo{journal}{Astropart. Phys.} \bibinfo{volume}{33}
  (\bibinfo{year}{2010}) \bibinfo{pages}{341--350}.
\bibitem[{{Engel} et~al.(2005){Engel}, {Kalmykov}, and
  {Konstantinov}}]{EngelKalmykovKonstantinovICRC2005}
\bibinfo{author}{R.~{Engel}}, \bibinfo{author}{N.~N. {Kalmykov}},
  \bibinfo{author}{A.~A. {Konstantinov}},
\newblock \bibinfo{title}{{Simulation Cherenkov and Synchrotron Radio Emission
  in EAS}},
\newblock in: \bibinfo{booktitle}{Proc. of the 29th ICRC, Pune, India},
  volume~\bibinfo{volume}{6}, pp. \bibinfo{pages}{9--12}.
\bibitem[{EGS(????)}]{EGSnrc}
\bibinfo{note}{{http://www.irs.inms.nrc.ca/EGSnrc/EGSnrc.html}}.
\bibitem[{{Werner} and {Scholten}(2008)}]{WernerScholten2008}
\bibinfo{author}{K.~{Werner}}, \bibinfo{author}{O.~{Scholten}},
\newblock \bibinfo{title}{{Macroscopic Treatment of Radio Emission from Cosmic
  Ray Air Showers based on Shower Simulations}},
\newblock \bibinfo{journal}{Astroparticle Physics} \bibinfo{volume}{29}
  (\bibinfo{year}{2008}) \bibinfo{pages}{393--411}.
\bibitem[{Alvarez-Mu\~niz et~al.(2010)Alvarez-Mu\~niz, Romero-Wolf, and
  Zas}]{AlvarezMunizRomeroWolfZas2010}
\bibinfo{author}{J.~Alvarez-Mu\~niz}, \bibinfo{author}{A.~Romero-Wolf},
  \bibinfo{author}{E.~Zas},
\newblock \bibinfo{title}{\ifmmode \check{C}\else \v{C}\fi{}erenkov radio
  pulses from electromagnetic showers in the time domain},
\newblock \bibinfo{journal}{Phys. Rev. D} \bibinfo{volume}{81}
  (\bibinfo{year}{2010}) \bibinfo{pages}{123009}.
\bibitem[{{James} et~al.(2010){James}, {Falcke}, {Huege}, and
  {Ludwig}}]{JamesFalckeHuege2010}
\bibinfo{author}{C.~W. {James}}, \bibinfo{author}{H.~{Falcke}},
  \bibinfo{author}{T.~{Huege}}, \bibinfo{author}{M.~{Ludwig}},
\newblock \bibinfo{title}{{An `endpoint' formulation for the calculation of
  electromagnetic radiation from charged particle motion}},
\newblock \bibinfo{journal}{Phys Rev.\ E} \bibinfo{volume}{submitted}
  (\bibinfo{year}{2010}). \bibinfo{note}{{arXiv:1007.4146}}.
\bibitem[{{de Vries} et~al.(2010){de Vries}, {van den Berg}, {Scholten}, and
  {Werner}}]{DeVriesScholtenWerner2010}
\bibinfo{author}{K.~D. {de Vries}}, \bibinfo{author}{A.~M. {van den Berg}}, \bibinfo{author}{O.~{Scholten}},
  \bibinfo{author}{K.~{Werner}},
\newblock \bibinfo{title}{{The Lateral Distribution Function of Coherent Radio
  Emission from Extensive Air Showers; Determining the Chemical Composition of
  Cosmic Rays}},
\newblock \bibinfo{journal}{Astropart. Phys.} \bibinfo{volume}{in press}
  (\bibinfo{year}{2010}). \bibinfo{note}{{arXiv:1008.3308}}.
\bibitem[{{Schoorlemmer} and {for the Pierre Auger
  Collaboration}(2010)}]{SchoorlemmerARENA2010}
\bibinfo{author}{H.~{Schoorlemmer}}, \bibinfo{author}{{for the Pierre Auger
  Collaboration}},
\newblock \bibinfo{title}{{Results from polarization studies of radio signals
  induced by cosmic rays at the Pierre Auger Observatory}},
\newblock \bibinfo{journal}{these proceedings}  (\bibinfo{year}{2010}).

\end{thebibliography}



\end{document}